
\input epsf
\input phyzzx
\overfullrule=0pt
\hsize=6.5truein
\vsize=9.0truein
\voffset=-0.1truein
\hoffset=-0.1truein
%
%
\def\BH{\hbox{\sevenrm bh}}
\def\half{{1\over 2}}
\def\inn{\hbox{\sevenrm in}}
\def\out{\hbox{\sevenrm out}}
\def\pa{\partial}
\def\scrip{{\cal I}^+}

\def\scripr{{\cal I}_R^+}

\def\scrimr{{\cal I}_R^-}

\def\sp{\sigma^+}
\def\sm{\sigma^-}
\def\Var{\hbox{Var}}
\def\xp{x^+}
\def\xm{x^-}
\def\yp{y^+}
\def\ym{y^-}

\rightline{SU-ITP-93-15}
\rightline{June 1993}
\rightline{hep-th/9306069}

\vfill

\title{The Stretched Horizon and Black Hole Complementarity}

\vfill

%
%
\author{Leonard Susskind,\foot{susskind@dormouse.stanford.edu}
L\'arus Thorlacius,\foot{larus@dormouse.stanford.edu}
and John Uglum\foot{john@dormouse.stanford.edu}}

\vfill

\address{Department of Physics \break Stanford University, Stanford,
CA
94305-4060}

\vfill

%
%
\abstract
{\singlespace
Three postulates asserting the validity of conventional quantum
theory, semi-classical general relativity and the statistical basis
for thermodynamics are introduced as a foundation for the study of
black hole evolution.  We explain how these postulates may be
implemented in a ``stretched horizon'' or membrane description of the
black hole, appropriate to a distant observer.  The technical
analysis is illustrated in the simplified context of 1+1 dimensional
dilaton gravity.  Our postulates imply that the dissipative
properties of the stretched horizon arise from a course graining of
microphysical degrees of freedom that the horizon must possess.  A
principle of black hole complementarity is advocated.  The overall
viewpoint is similar to that pioneered by 't~Hooft but the detailed
implementation is different.}

%
%
PACS categories: 04.60.+n, 12.25.+e, 97.60.Lf
\vfill\endpage

%

\REF\hawk{S.~W.~Hawking
\journal Comm .Math. Phys. & 43 (75) 199
\journal Phys. Rev. & D14 (76) 2460.}

\REF\dpage{For an early critique of Hawking's proposal, see
D.~N.~Page
\journal Phys. Rev. Lett. & 44 (80) 301.}

\REF\tHooftone{G.~'t~Hooft
\journal Nucl. Phys. & B335 (90) 138, and references therein.}

\REF\svv{E.~Verlinde and H.~Verlinde,
{\it A Unitary S-matrix for 2D Black Hole Formation and Evaporation,}
Princeton preprint, PUPT-1380,IASSNS-HEP-93/8, February
1993;\hfil\break
K. Schoutens, E.~Verlinde, and H.~Verlinde,
{\it Quantum Black Hole Evaporation,} Princeton preprint,
PUPT-1395,IASSNS-HEP-93/25, February 1993.}

\REF\Bek{J.~D.~Bekenstein
\journal Phys. Rev. & D9 (74) 3292.}


\REF\hrhz{R.~S.~Hanni and R.~Ruffini
\journal Phys. Rev. & D8 (73) 3259; \hfil\break
P.~Hajicek
\journal Comm. Math. Phys. & 36 (74) 305; \hfil\break
R.~L.~Znajek
\journal Mon. Not. Roy. Astron. Soc. & 185 (78) 833.}

\REF\damzna{T.~Damour
\journal Phys. Rev. & D18 (78) 3598; {\it Proceedings of the Second
Marcel Grossmann Meeting on General Relativity,} ed. R.~Ruffini, p.
587, (North-Holland, Amsterdam) 1982;\hfil\break
R.~L.~Znajek,
{\it ``Black Holes, Apparent Horizons, Entropy, and Temperature,''}
University of Cambridge manuscript, 1981, unpublished; {\it ``A
Hartle-Hawking Formula for Apparent Horizons,''} University of
Cambridge manuscript, 1981, unpublished.}

\REF\thornes{K.~S.~Thorne and D.~A.~MacDonald
\journal Mon. Not. Roy. Astron. Soc. & 198 (82) 339;\hfil\break
W.~H.~Zurek and K.~S.~Thorne
\journal Phys. Rev. Lett. & 54 (85) 2171; \hfil\break
R.~H.~Price and K.~S.~Thorne
\journal Phys. Rev. & D33 (86) 915.}

\REF\Mempar{K.~S.~Thorne, R.~H.~Price, and D.~A.~MacDonald, P{\it
Black
Holes:  The Membrane Paradigm},  Yale University Press, 1986, and
references therein.}

\REF\tHoofttwo{G.~'t~Hooft, private communication.}

\REF\CGHS{C.~G.~Callan, S.~B.~Giddings, J.~A.~Harvey and
A.~Strominger
\journal Phys. Rev. & D45 (92) R1005.}

\REF\dilbh{S.~B.~Giddings and A.~Strominger
\journal Phys. Rev. & D46 (92) 627.}

\REF\bddo{T.~Banks, A.~Dabholkar, M.~R.~Douglas and M.~O'Loughlin
\journal Phys. Rev. & D45 (92) 3607.}

\REF\RSTone{J.~G.~Russo, L.~Susskind, and L.~Thorlacius
\journal Phys. Lett. & B292 (92) 13.}

\REF\bghs{B.~Birnir, S.~B.~Giddings, J.~A.~Harvey and A.~Strominger
\journal Phys. Rev. & D46 (92) 638.}

\REF\haw{S.~W.~Hawking
\journal Phys. Rev. Lett. & 69 (92) 406.}

\REF\lslt{L.~Susskind and L.~Thorlacius,
\journal Nucl. Phys. & B382 (92) 123.}

\REF\Stromone{A. ~Strominger
 \journal Phys. Rev. & D46 (92) 4396.}

\REF\jrat{J.~G.~Russo and A.~A.~Tseytlin
\journal Nucl. Phys. & B382 (92) 259;\hfil\break
T.~Burwick, A.~Chamseddine
\journal Nucl. Phys. & B384 (92) 411.}

\REF\bilcal{A.~Bilal and C.~G.~Callan
\journal Nucl. Phys. & B394 (93) 73.}

\REF\dealw{S.~P.~de~Alwis
\journal Phys. Lett. & B289 (92) 278
\journal Phys. Lett. & B300 (93) 330
\journal Phys. Rev. & D46 (92) 5429.}

\REF\gidstr{S.~B.~Giddings and A.~Strominger
\journal Phys. Rev. & D47 (93) 2454.}

\REF\RSTtwo{J.~G.~Russo, L.~Susskind, and L.~Thorlacius
\journal Phys. Rev. & D46 (92) 3444.}

\REF\RSTthree{J.~G.~Russo, L.~Susskind, and L.~Thorlacius
\journal Phys. Rev. & D47 (93) 533.}

\REF\past{Y.~Park and A.~Strominger
\journal Phys. Rev. & D47 (93) 1569.}

\REF\lowe{D.~Lowe
\journal Phys. Rev. & D47 (93) 2446;\hfil\break
S.~W.~Hawking and J.~M.~Stewart, {\it Naked and Thunderbolt
Singularities in Black Hole Evaporation,} University of Cambridge
preprint, PRINT-92-0362, July 1992;\hfil\break
T.~Piran and A.~Strominger, {\it Numerical Analysis of Black Hole
Evaporation,} UCSB preprint, NSF-ITP-93-36, April 1993.}

\REF\page{D.~N.~Page, {\it Black Hole Information,} University of
Alberta preprint, Alberta-Thy-23-93, May 1993, and references
therein.}

\REF\ssstt{N.~Seiberg, S.~Shenker, L.~Susskind, L.~Thorlacius, and
J.~Tuttle, work in progress.}

\REF\seishe{N.~Seiberg and S.~Shenker, private communication.}

\REF\bobl{R.~B.~Laughlin, private communication.}

%
%

%
%
\chapter{Introduction}

The formation and evaporation of a macroscopic black hole is a
complex process which certainly leads to a practical loss of
information and an increase of thermal entropy.  The same is true of
almost all macroscopic phenomena.  It is exceedingly difficult to
keep track of all the degrees of freedom involved when a large block
of ice melts or a bomb explodes, but {\it in principle\/} it can be
done.  According to the standard rules of quantum field theory in a
fixed Minkowski spacetime, the time evolution of any system from a
given initial state is described unambiguously by a unitary
transformation acting on that state, and in this sense there is never
any loss of fundamental, {\it fine grained\/} information.

The situation is less clear when gravitational effects are taken into
account.  It has been suggested [\hawk] that fundamental information
about the quantum state of matter undergoing gravitational collapse
will be irretrievably lost behind the event horizon of the resulting
black hole.  In this view, the Hawking emission from the black hole
is in the form of thermal radiation, which carries little or no
information about the initial quantum state of the system.  If the
black hole evaporates completely, that information would be lost, in
violation of the rules of quantum theory.  We believe such a
conclusion is unnecessary [\dpage].

This paper is based on the assumption that black hole evolution can
be reconciled with quantum theory,  a viewpoint which has been most
strongly advocated by 't~Hooft~[\tHooftone].\foot{This viewpoint has
more recently also been put forward in the work of K.~Schoutens,
E.~Verlinde, and H.~Verlinde, in the context of a two-dimensional toy
model [\svv].}  We shall introduce three postulates upon which we
believe a phenomenological description of black holes should be
based.  These postulates extrapolate the validity of the empirically
well-established principles of quantum theory, general relativity,
and statistical mechanics to phenomena involving event horizons.  We
argue that a phenomenological description of black holes, based on
the idea of a ``stretched horizon''\foot{The definition of the
stretched horizon will be given in Section~3.} which can absorb,
thermalize, and re-emit information, is consistent with these
postulates.

The postulates are the following:

$\bullet$ {\bf Postulate 1:}  {\it The process of formation and
evaporation of a black hole, as viewed by a distant observer, can be
described entirely within the context of standard quantum theory.  In
particular, there exists a unitary $S-$matrix which describes the
evolution from infalling matter to outgoing Hawking-like radiation.}

This postulate agrees with the $S$-matrix approach of 't~Hooft
[\tHooftone].  Furthermore, we assume there exists a Hamiltonian
which generates the evolution for finite times.

The second postulate states the validity of semi-classical
gravitation theory,
including quantum corrections to the classical equations of motion,
in the region outside a massive black hole.  The semi-classical
equations should contain enough quantum corrections to account for
the outgoing Hawking flux and the evaporation of the black hole.

$\bullet$ {\bf Postulate 2:}  {\it Outside the stretched horizon of a
massive black hole, physics can be described to good approximation by
a set of semi-classical field equations.}

No consistent formulation of such a set of equations has been
achieved in four-dimensional gravity.  Furthermore, the concept of a
dynamical stretched horizon is quite complicated for arbitrary
time-dependent black holes in four dimensions.   The situation is
much simpler in two-dimensional gravity and recent months have seen
significant progress in constructing a semi-classical description
appropriate for Postulate~2.  The stretched horizon is easily defined
in this simplified context.  For these reasons we shall illustrate
the stretched horizon idea using a two-dimensional toy model.   The
semi-classical equations, whose nature is partly field theoretic and
partly thermodynamic, describe the average energy flow and evolution
of the horizon.

The third postulate is concerned with the validity of black hole
thermodynamics and its connection with Postulate 1.  Specifically, we
assume that the origin of the thermodynamic behavior of the black
hole is the coarse graining of a large, complex, ergodic, but
conventionally quantum mechanical system.

$\bullet$ {\bf Postulate 3:}  {\it To a distant observer, a black
hole appears to be a quantum system with discrete energy levels.  The
dimension of the subspace of states describing a black hole of mass
$M$ is the exponential of the Bekenstein entropy $S(M)$ [\Bek].}

In particular, we assume there is no infinite additive constant in
the entropy.

The above three postulates all refer to observations performed from
outside the
black hole.  Although we shall not introduce specific postulates
about observers who fall through the global event horizon, there is a
widespread belief which we fully share.  The belief is based on the
equivalence principle and the fact that the global event horizon of a
very massive black hole does not have large curvature, energy
density, pressure, or any other invariant signal of its presence.
For this reason, it seems certain that a freely falling observer
experiences nothing out of the ordinary when crossing the horizon.
It is this assumption which, upon reflection, seems to be sharply at
odds with Postulate 1.  Let us review the argument.

Consider a Penrose diagram for the formation and evaporation of a
black hole, as in Figure~1.  Foliate the spacetime with a family of
space-like Cauchy surfaces, as shown.  Some of the Cauchy surfaces
will lie partly within the black hole.  Consider the surface
$\Sigma_P$ which contains the point $P$ where the global event
horizon intersects the curvature singularity.  $P$ partitions
$\Sigma_P$ into two disjoint surfaces $\Sigma_{\BH}$ and
$\Sigma_{\out}$ which lie inside and outside the black hole,
respectively.
\vskip 15pt
\vbox{
{\centerline{\epsfsize=3.0in \hskip 2cm \epsfbox{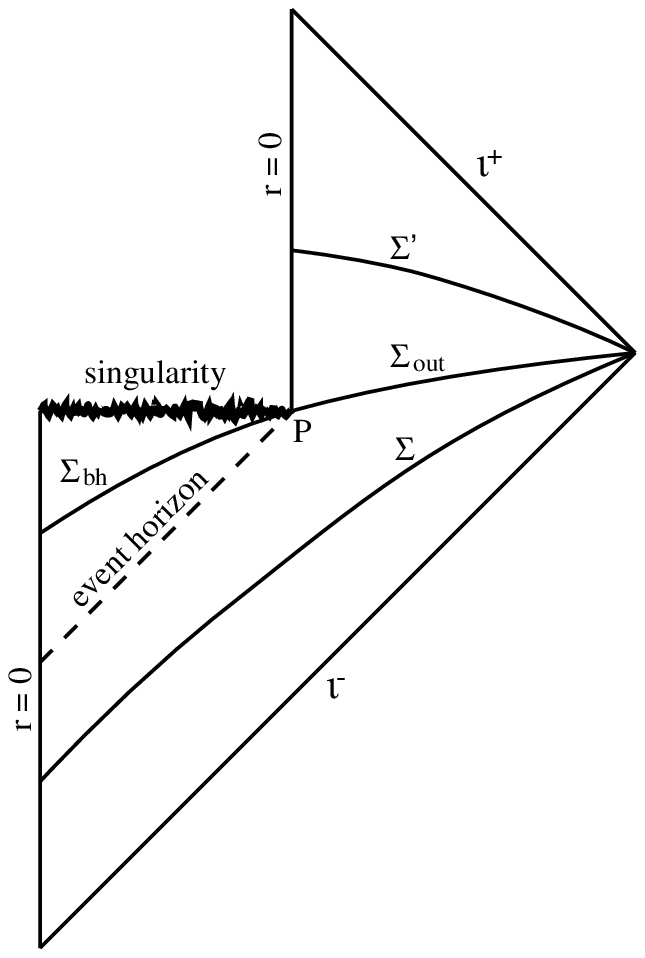}}}
\vskip 12pt
{\centerline{\tenrm FIGURE 1. Penrose diagram for black hole
evolution.}}
\vskip 15pt}

Now assume that there exists a linear Schr\"odinger equation,
derivable from a local quantum field theory, which describes the
evolution of state vectors from one Cauchy surface to the next.  An
initial state $\ket{\Psi(\Sigma)}$ defined on some Cauchy surface
$\Sigma$ which does not intersect the black hole can be evolved
without encountering any singularity until the surface $\Sigma_P$ is
reached.  On $\Sigma_P$ the Hilbert space of states $\cal H$ can be
written as a tensor product space ${\cal H} = {\cal H}_{\BH} \otimes
{\cal H}_{\out}$ of functionals of the fields on $\Sigma_{\BH}$ and
$\Sigma_{\out}$, respectively.

Next, consider evolving the state further to some surface $\Sigma'$
in the
future, as indicated in Figure~1.  The resulting state,
$\ket{\Psi(\Sigma')}$, represents the observable world long after the
black hole has evaporated.  According to Postulate 1,
$\ket{\Psi(\Sigma')}$ must be a {\it pure state} which is related to
the original incoming state $\ket{\Psi(\Sigma)}$ by a linear operator
$S$, the $S-$matrix.  By assumption, $\ket{\Psi(\Sigma')}$ has
evolved by the Schr\"odinger equation from some state
$\ket{\chi(\Sigma_{\out})}$ defined on $\Sigma_{\out}$, which must
then also be a pure state.  This, in turn, implies that
$\ket{\Psi(\Sigma_P)}$ must be a product state,
$$
\ket{\Psi(\Sigma_P)} = \ket{\Phi(\Sigma_{\BH})} \otimes
\ket{\chi(\Sigma_{\out})} \>,
\eqn\oneone
$$
where $\ket{\Phi(\Sigma_{\BH})} \in {\cal H}_{\BH}$ and
$\ket{\chi(\Sigma_{\out})} \in {\cal H}_{\out}$.  The product state
is obtained by linear Schr\"odinger evolution from the initial state
$\ket{\Psi(\Sigma)}$, but as seen above, the external factor
$\ket{\chi(\Sigma_{\out})}$ alone depends linearly on
$\ket{\Psi(\Sigma)}$, so we arrive at the conclusion that the state
inside the black hole, $\ket{\Phi(\Sigma_{\BH})}$, must be
independent of the initial state.  In other words, all distinctions
between initial states of infalling matter must be obliterated before
the state crosses the global event horizon. But this is an entirely
unreasonable violation of the equivalence principle. Therefore, the
argument goes, the outside observer cannot see a pure state.

Although this conclusion seems to follow from fairly general
principles, we believe it is unwarranted.  The assumption of a state
$\ket{\Psi(\Sigma_P)}$ which simultaneously describes both the
interior and the exterior of a black hole seems suspiciously
unphysical.  Such a state can describe correlations which have no
operational meaning, since an observer who passes behind the event
horizon can never communicate the result of any experiment performed
inside the black hole to an observer outside the black hole.  The
above description of the state lying in the tensor product space
${\cal H}_{\BH} \otimes {\cal H}_{\out}$ can only be made use of by a
``superobserver'' outside our universe.  As long as we do not
postulate such observers, we see no logical contradiction in assuming
that a distant observer sees all infalling information returned in
Hawking-like radiation, and that the infalling observer experiences
nothing unusual before or during horizon crossing. Only when we try
to give a combined description, with a standard quantum theory valid
for both observers, do we encounter trouble.  Of course, it may be
argued that a quantum field theoretic description of gravity dictates
just such a description, whether we like it or not.  If this is the
case, such a quantum field theory is inconsistent with our
postulates; therefore, one or the other is incorrect.

Let us now consider the process of formation and evaporation of a
black hole as seen by a distant observer.  It is well known that the
physics of a {\it classical\/}, quasistationary black hole can be
described by outside observers in terms of a ``stretched horizon'',
which behaves in all respects like a physical membrane with certain
mechanical, electrical, and thermal properties [\hrhz-\Mempar].  The
description is coarse-grained in character, by which we mean that it
has the typical time irreversibility and dissipative properties of a
system described by ordinary thermodynamics.

The membrane is very real to an outside observer.  For example, if
such an observer is suspended just above the stretched horizon, he or
she will observe an intense flux of energetic radiation apparently
emanating from the membrane.  If provided with an electrical
multimeter, our observer will discover that the membrane has a
surface resistivity of 377 ohms.  If disturbed, the stretched horizon
will respond like a viscous fluid, albeit with negative bulk
viscosity.  And finally, the observed entropy of the massive black
hole is proportional to the area of the stretched horizon.  If, on
the other hand, the observer attempts to determine if the membrane is
real by letting go of the suspension mechanism and falling freely
past the stretched horizon, the membrane will disappear.  However,
there is no way to report the membrane's lack of substance to the
outside world.  In this sense, there is {\it complementarity\/}
between observations made by infalling observers who cross the event
horizon and those made by distant observers.\foot{A similar view has
been expressed by 't~Hooft~[\tHoofttwo].}

We believe that Postulates 1-3 are most naturally implemented by
assuming that the coarse grained thermodynamic description of an
appropriately defined stretched horizon has an underlying
microphysical basis.  In other words, from the point of view of an
outside observer, {\it the stretched horizon is a boundary surface
equipped with microphysical degrees of freedom that appear in the
quantum Hamiltonian used to describe the observable world}.  These
degrees of freedom must be of sufficient complexity that they behave
ergodically and lead to a coarse-grained, dissipative description of
the membrane.

Much of this paper is concerned with the illustration of the concept
of the stretched horizon in the context of two-dimensional dilaton
gravity, for which a semi-classical description has been formulated
[\CGHS-\lowe].  We review this formalism in section~2.  In section~3
we define the stretched horizon and study its behavior and
kinematics.  The definition of the stretched horizon which we find
most useful differs somewhat from that used for classical black holes
in [\Mempar].  Our semi-classical stretched horizon is minimally
stretched, in that its area is only one Planck unit larger than the
area of the global event horizon itself, whereas in [\Mempar], the
areas of the two horizons differ by a macroscopic amount.  The
evolution of the stretched horizon can be followed throughout the
entire process of black hole formation and evaporation, except for
the final period when the black hole is of Planckian size.  In
section~4, we show that the stretched horizon has statistical
fluctuations which cause its area to undergo brownian motion, and to
diffuse away from its classical evolution.  The semi-classical theory
does not provide a microphysical description, but it helps in
formulating a kinematic framework for one.  In section~5 we examine
consequences of the postulates.

Our assumptions have as consequences certain broad features of the
way information is stored in the approximately thermal Hawking
radiation.  The information is not returned slowly in far infrared
quanta long after most of the infalling energy has been re-radiated.
Nor is it stored in stable light remnants.  It is instead found in
long-time, non-thermal correlations between quanta emitted at very
different times, as advocated by Don Page [\page].  The viewpoint of
this paper is essentially that of 't~Hooft [\tHooftone].  However, we
believe that the stretched horizon is a very complex and chaotic
system.  Even if the microscopic laws were known, computing an
$S-$matrix [\tHooftone,\svv] would, according to this view, be as
daunting as computing the scattering of laser light from a chunk of
black coal.  The validity of quantum field theory in this case is not
assured by exhibiting an $S-$matrix, but by identifying the
underlying atomic structure and constructing a Schr\"odinger equation
for the many particles composing the coal and the photon field to
which it is coupled.  Although the equations cannot be solved, we
nevertheless think we understand the route from quantum theory to
apparently thermal radiation via statistical mechanics.  In the case
of the stretched horizon, the underlying microphysics is not yet
understood, but we hope that that the semi-classical considerations
in this paper will help in identifying the appropriate degrees of
freedom.

%
%
\chapter{Two-dimensional dilaton gravity}

It is very useful to have a simplified setting in which to study
black hole physics.  Callan, Giddings, Harvey and Strominger (CGHS)
suggested for this purpose two-dimensional dilaton gravity coupled to
conformal matter [\CGHS]. Their model can be exactly solved at the
classical level and has solutions which are two-dimensional analogs
of black holes.  Quantum corrections are much more amenable to study
in this theory than in four-dimensional Einstein gravity.  In this
section we will review the classical theory and then show how quantum
corrections can be implemented via a set of semi-classical equations
which can be solved explicitly.  This material is not new but it
serves to fix notation and makes our discussion for the most part
self-contained.

\section{Classical theory}

The classical CGHS-model of two-dimensional dilaton gravity is
defined by the action functional
$$
S_0[f_i, \phi, g] =
{1 \over 2\pi} \int d^2x \sqrt{-g}[e^{-2\phi}
(R + 4(\nabla \phi)^2 + 4\lambda^2)
- \half \sum_{i=1}^N (\nabla f_i)^2] \>.
\eqn\scghs
$$
It can be viewed as an effective action for radial modes of
near-extreme, magnetically charged black holes in four-dimensional
dilaton gravity [\CGHS-\bddo].  The two-dimensional length scale
$\lambda$ is inversely related to the magnetic charge of the
four-dimensional black hole.  For convenience, we shall choose units
in which $\lambda = 1$.  In the region of the four-dimensional
geometry where the two-dimensional effective description applies, the
physical radius of the local transverse two-sphere is governed by the
dilaton field, $r(x^0,x^1)=e^{-\phi(x^0,x^1)}$.  The area calculated
from this radius is proportional to the Bekenstein entropy of the
four-dimensional black hole and accordingly we will refer to the
function,
$$
{\cal A} = e^{-2\phi} \>,
\eqn\areaone
$$
as the classical ``area'' function in the two-dimensional effective
theory.

The classical equations of motion are
$$
2 \nabla_\mu \nabla_\nu \phi - 2 g_{\mu \nu}
(\nabla^2 \phi - (\nabla \phi)^2 + 1)
- e^{2 \phi}\,T_{\mu \nu} = 0 \>,
\eqn\Tmnone
$$
$$
{1\over 4}R + \nabla^2 \phi
- (\nabla \phi)^2 + 1 = 0 \>,
\eqn\dilone
$$
$$
\nabla^2 f_i = 0  \>,
\eqn\feqone
$$
where $T_{\mu \nu}$ is the matter energy-momentum tensor, given by
$$
T_{\mu \nu} =
\half \sum_{i=1}^N [\nabla_\mu f_i\nabla_\nu f_i
- \half g_{\mu \nu} (\nabla f_i)^2] \>.
\eqn\Tmntwo
$$

To solve the above equations we go to conformal gauge and choose
light-cone
coordinates $(\xp,\xm)$ in which the line element is $ds^2 = -e^{2
\rho}\, d\xp \, d\xm$.  The equations of motion are then
$$
2\pa_{\pm}^2 \phi
- 4\pa_{\pm} \phi\pa_{\pm} \rho
- e^{2 \phi}\, T_{\pm \pm} = 0  \>,
\eqn\elone
$$
$$
 4\pa_+ \phi\pa_- \phi
- 2 \pa_+ \pa_- \phi + e^{2 \rho} = 0  \>,
\eqn\eltwo
$$
$$
2 \pa_+ \pa_- \rho - 4 \pa_+ \pa_- \phi
+ 4\pa_+ \phi\pa_- \phi
+ e^{2 \rho} = 0  \>,
\eqn\elthree
$$
$$
\pa_+ \pa_- f_i = 0  \>,
\eqn\elfour
$$
and the non-vanishing components of the matter energy-momentum tensor
are given by
$$
T_{\pm \pm} = \half \sum_{i=1}^N (\pa_{\pm} f_i)^2  \>.
\eqn\Tmnthree
$$
The action \scghs\ written in conformal gauge, has a global symmetry
generated by the conserved current $j^{\mu} = \nabla^{\mu} (\rho -
\phi)$, and thus
$$
\pa_+ \pa_- (\rho - \phi) = 0  \>.
\eqn\gaugefix
$$
This equation allows one to fix the remaining subgroup of conformal
transformations by choosing coordinates in which $\rho = \phi.$  We
will denote any set of light-cone coordinates in which $\rho = \phi$
as {\it Kruskal coordinates.}

The vacuum solution is given by
$$
\eqalign{f_i &= 0\>, \cr
e^{-2 \phi} = e^{-2\rho} &= -\yp \ym\>. \cr}
\eqn\ldvone
$$
If we define new coordinates $\sigma^{\pm}$ by the transformation
$y^{\pm}=\pm e^{\pm \sigma^{\pm}}$, we find that the spacetime can be
identified as two-dimensional Minkowski space, with line element
$ds^2=-d\sigma^+\,d\sigma^-$.  In these coordinates, the dilaton
field is given by
$$
\phi = -\half (\sigma^+ - \sigma^-) \equiv -\sigma  \>,
\eqn\ldvtwo
$$
and thus this solution is called the {\it linear dilaton vacuum.}

\section{Classical black holes}

A black hole is defined as a region of spacetime which does not lie
in the causal past of future null infinity $\scrip$, \ie\ light rays
which have their origin inside the black hole can never escape to
$\scrip$.  The global event horizon, denoted $H_G$, is the boundary
of the black hole region.  It is a null surface representing the last
light rays which are trapped by the black hole.  It is important to
note that the definitions of the black hole region and global event
horizon are not local.  To define a black hole and its global event
horizon one must have knowledge of the entire spacetime manifold - in
particular, one must be able to find the causal past of $\scrip$.  As
a result, observers will not be able to tell when they pass through
the global event horizon of a massive black hole.

The linear dilaton vacuum solution \ldvone\ can easily be generalized
to a one-parameter family of static black hole solutions,
$$
\eqalign{f_i &= 0\>, \cr e^{-2 \phi}
= e^{-2\rho} &= M_0 - \yp \ym\>, \cr}
\eqn\ebhone
$$
where $M_0>0$ is proportional to the Arnowitt-Deser-Misner (ADM) mass
of the black hole.  The scalar curvature is given by
$$
R = {{4M_0} \over {M_0 - \yp \ym}} \>,
\eqn\ebhtwo
$$
which becomes infinite when $M_0 - \yp \ym = 0$.  Thus there are two
curvature singularities, which asymptotically approach the null
curves $y^{\pm} = 0$.  The Penrose diagram for this solution is
displayed in Figure~2.  One of the curvature singularities does not
lie in the causal future of any point of the spacetime and is the
singularity of a white hole.  The other, of course, is the black hole
singularity.

\vskip 15pt
\vbox{
{\centerline{\epsfsize=3.0in \hskip 2cm \epsfbox{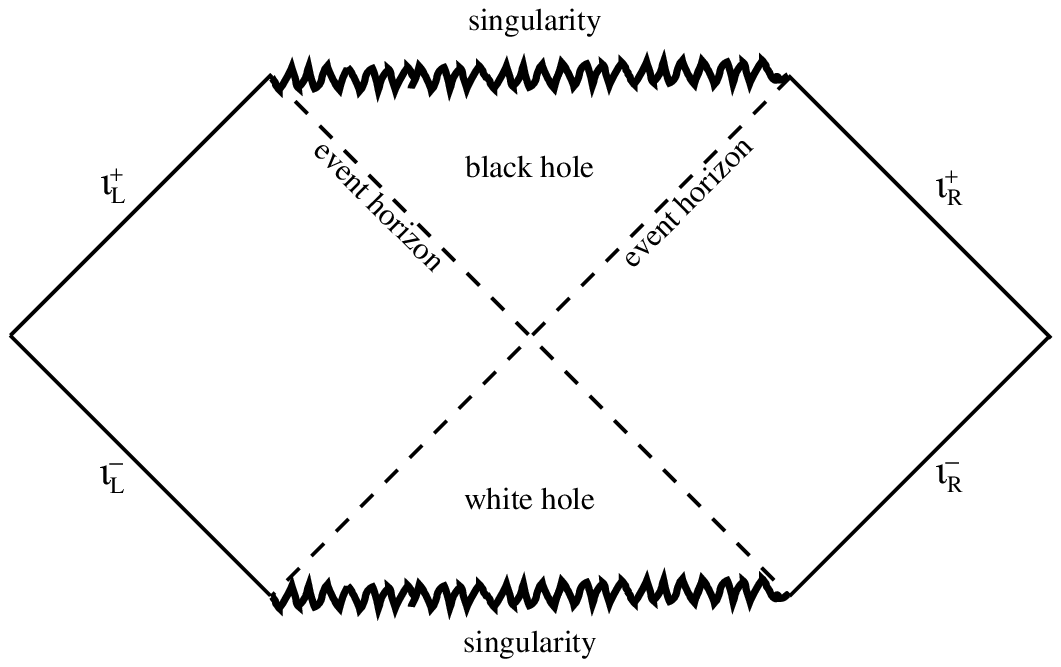}}}
\vskip 12pt
{\centerline{\tenrm FIGURE 2. Penrose diagram for the eternal black
hole solution.}}
\vskip 15pt}

A more physically interesting set of solutions describes black hole
formation by incoming matter,
$$
\eqalign{f_i &= f^+_i(\yp)\>, \cr
e^{-2 \phi} = e^{-2\rho} &= M(\yp)
-\yp \bigl(\ym + P_+(\yp) - P_{\infty}\bigr)\>, \cr}
\eqn\imone
$$
where $M(\yp)$ and $P_+(\yp)$ are the following functions of the
infalling matter:
$$
M(\yp) = \int_0^{y^+} du\, u\, T_{++}(u) \>,
\qquad
P_+(\yp) = \int_0^{\yp} du\, T_{++}(u)  \>,
\eqn\massone
$$
and $P_{\infty} = P_+(\yp{=}\infty)$.  The scalar curvature is
$$
R = {{4M(\yp)} \over {M(\yp)
- \yp (\ym + P_+(\yp) - P_{\infty}) }} \>.
\eqn\imtwo
$$
The functions $f^+_i$ are taken to be non-vanishing only on the
interval $[\yp_1, \yp_2]$, \ie\ the matter flux is switched on for a
finite time interval.  For $y^+<y_1^+$, the solution reduces to the
linear dilaton vacuum, \ldvone , with $y^-$ shifted by $P_{\infty}$,
and for $\yp >\yp_2$, the solution is an eternal black hole solution
described by \ebhone\ with $M_0$ replaced by $M_{\infty}=
M(\yp{=}\infty)$.  The Penrose diagram is shown in Figure~3.  The
global event horizon $H_G$ is the curve $\ym = 0$.

\vskip 15pt
\vbox{
{\centerline{\epsfsize=3.0in \hskip 2cm \epsfbox{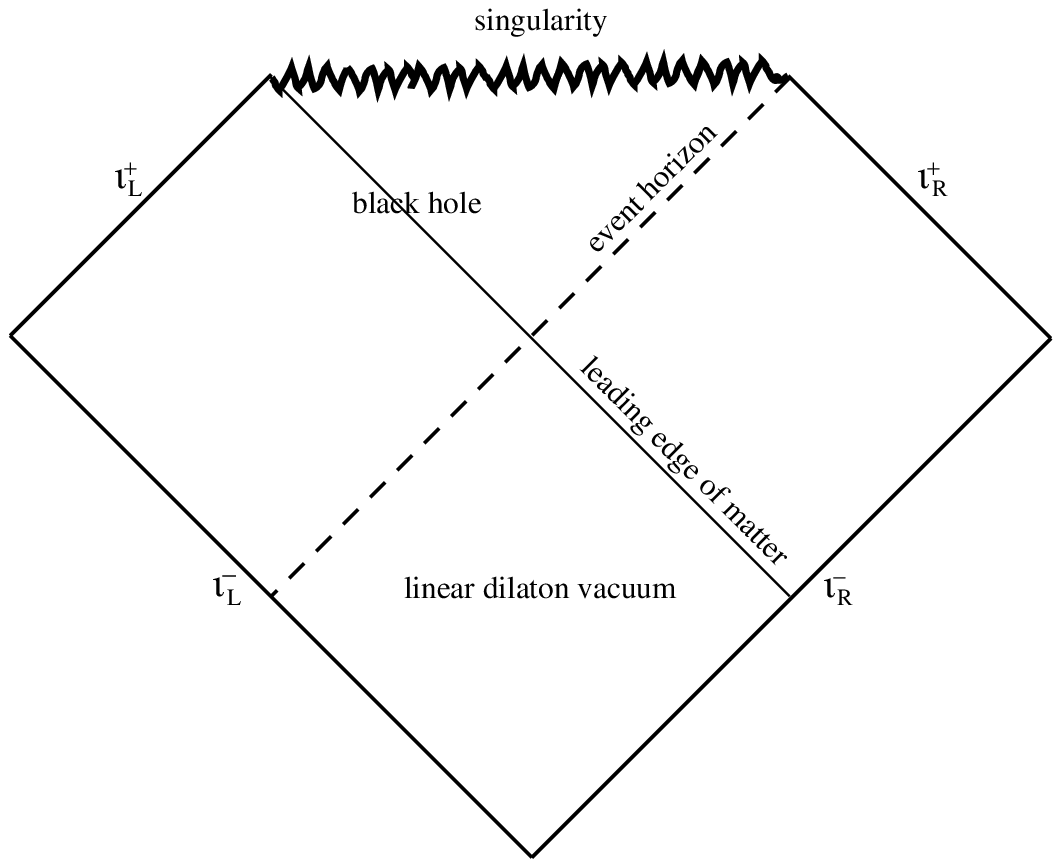}}}
\vskip 12pt
{\centerline{\tenrm FIGURE 3. Penrose diagram for the infall
solution.}}
\vskip 15pt}

Kruskal coordinates are not convenient for the description of
processes by an external observer.  One would like to find a
coordinate system which covers only the region exterior to the black
hole, and reduces to Minkowski coordinates far from the black hole,
so that physical quantities can be defined unambiguously.  We define
{\it tortoise coordinates\/} $(t,\sigma)$ as
$$
\eqalign{t &= \half \log (-{\yp\over \ym}) \>, \cr
\sigma &= \half \log (- \yp \ym) \>. \cr}
\eqn\tortone
$$
The line element of the gravitational collapse solution, \imone ,
takes the form
$$
ds^2 = \Lambda(t,\sigma)\> \bigl[ -dt^2 + d\sigma^2 \bigr] \>,
\eqn\torttwo
$$
where
$$
\Lambda(t,\sigma) =
\bigl[1+M(t,\sigma)\,e^{-2\sigma}
-(P_+(t,\sigma)-P_{\infty})\,e^{(t-\sigma)}
\bigr]^{-1}\>.
\eqn\lambdaone
$$
The tortoise coordinates are asymptotically flat, and the line
element is conformal to that of Minkowski space.  The global event
horizon is at $t = \infty$, $\sigma = - \infty$, and for the eternal
black hole solution, $({\pa\over \pa t})$ is a time-like Killing
vector.  The light-cone coordinates $\sigma^{\pm} = t \pm \sigma$
exactly cover $\scrimr$ and $\scripr$, respectively, so we see that
tortoise coordinates are the coordinates appropriate for the
description of processes as seen by asymptotic inertial observers.
They provide a time variable which covers the entire region
accessible to an outside observer and we assume the existence of a
Hamiltonian, which generates translations of this time variable.

\section{Semi-classical theory}

Our second postulate assumes that a semi-classical approximation to
gravitation theory can be developed systematically.  In the
simplified world of two-dimensional dilaton gravity this can be
achieved by the addition of certain quantum corrections to the
classical equations of motion, as first described in the
groundbreaking work of Callan {\it et al.} [\CGHS].  These
corrections arise from the conformal anomaly of the matter fields in
the theory, which takes the form
$$
\vev{T_\mu^{\phantom{\mu}\mu}} = {N \over 24} R \>.
\eqn\twoone
$$
The semi-classical CGHS model is obtained by adding to the classical
action, \scghs , the associated Liouville term,
$$
S_L = - {N \over 96\pi} \int
d^2x \sqrt{-g(x)} \int d^2x' \sqrt{-g(x')}
\> R(x) G(x;x') R(x') \>,
\eqn\louie
$$
where $G$ is a Green function for the operator $\nabla^2$.  This
incorporates the leading-order quantum back-reaction on the geometry
due to the matter fields.  The original CGHS-equations have not been
solved in closed form (see [\lowe] for results of numerical studies)
but subsequent work led to a set of semi-classical equations which
can be solved exactly [\bilcal-\RSTtwo].  In the following we will
use the model introduced by Russo, Susskind and Thorlacius (RST) and
give a summary of the results of [\RSTtwo,\RSTthree].  This model is
obtained by including in the effective action a local covariant
counterterm,
$$
-{N \over 48\pi} \int
d^2x \sqrt{-g}\,\phi\,R \>,
\eqn\cterm
$$
in addition to the non-local Liouville term.  This turns out to
simplify the analysis and physical interpretation of the
semi-classical solutions.

We work in conformal gauge and use light-cone coordinates
$(\yp,\ym)$.  The effective action becomes
$$\eqalign{
S_{eff} = {1 \over \pi} \int d^2y
\bigl\{ e^{-2 \phi}
[2\pa_+&\pa_-\rho-4\pa_+\phi\pa_-\phi+e^{2 \rho}]
+ \half \sum_{i=1}^N \pa_+f_i \pa_-f_i  \cr
&- \kappa [\pa_+\rho\pa_-\rho + \phi\pa_+\pa_-\rho]
\bigr\}  \>, \cr}
\eqn\sefftwo
$$
where $\kappa = {N\over 12}$\foot{We will not go into the technical
issues involving reparametrization ghosts {\it etc.} which are
involved in the determination of the value of $\kappa$.  Our goal
here is limited to obtaining exactly solvable equations, which
incorporate the leading semi-classical corrections and exhibit
reasonable physical behavior, such as having a rate of Hawking
radiation proportional to the number of matter fields.}  The
constraint equations, which follow from varying $g_{\pm\pm}$, are
$$
(e^{-2\phi} + {\kappa \over 4})
[2\pa^2_{\pm}\phi - 4 \pa_{\pm}\rho \pa_{\pm}\phi]
- \kappa (\pa^2_{\pm}\rho - (\pa_{\pm} \rho)^2 - t_{\pm})
- T_{\pm\pm}  = 0 \>.
\eqn\conone
$$
Here $T_{\pm\pm}$ is the physical, observable flux of
energy-momentum.  There are subtleties involved in the regularization
of the composite operator.  We define $T_{\pm\pm}$ to be normal
ordered with respect to the asymptotically minkowskian tortoise
coordinates \tortone .

The functions $t_{\pm}(y^{\pm})$ reflect both the non-local nature of
the anomaly and the choice of boundary conditions satisfied by the
Green function $G$.  They are fixed by physical boundary conditions
on the semi-classical solutions.

If we define the two-component vector
$$
\Phi = \pmatrix{\phi \cr \rho \cr} \>,
\eqn\twotwo
$$
then the kinetic terms in the action \sefftwo\ may be written $(\pa_+
\Phi) \cdot M \cdot (\pa_- \Phi)$, and one finds that
$$
(-{\det (M)\over 4})^{-{1\over 4}}
= (e^{-2 \phi} -{\kappa\over 4})^{-\half}
\eqn\twothree
$$
plays the role of the gravitational coupling constant for the $f_i$
fields.  This coupling becomes infinite on a curve $\gamma_{cr}$ on
which the classical area function \areaone\ takes on the value
$$
{\cal A}_{cr} = {\kappa \over 4} \>.
\eqn\phicrit
$$
The curve $\gamma_{cr}$ has been interpreted to be a boundary of the
semi-classical spacetime [\lslt,\RSTthree], which plays the same role
as the surface $r = 0$ in the Schwarzschild solution of
four-dimensional Einstein gravity.  Accordingly, we have to impose a
boundary condition on $\gamma_{cr}$ when it is timelike.  Following
[\RSTthree], the boundary condition we use is to require the scalar
curvature to be finite on $\gamma_{cr}$.  This boundary condition
implements a weak form of the cosmic censorship hypothesis, in that
curvature singularities on $\gamma_{cr}$ will necessarily be
spacelike and cloaked by a global event horizon, except possibly for
isolated points.

We next define the fields\foot{Note that the normalizations of the
fields $\Omega$ and $\chi$ defined here differ by a factor of
$\sqrt{\kappa}$ from those given in [\RSTtwo,\RSTthree].}
$$\eqalign{
\Omega &= e^{-2 \phi} + {\kappa \over 2} \phi \>, \cr
\chi &= e^{-2 \phi} + \kappa (\rho - \half \phi) \>, \cr}
\eqn\omegatwo
$$
for which the effective action takes the simple form
$$
S_{eff} = {1 \over \pi} \int  d^2y
\bigl\{ {1 \over \kappa}
[-\pa_+\chi \pa_-\chi + \pa_+\Omega\pa_-\Omega ]
+ e^{{2 \over \kappa}(\chi - \Omega)}
+ \half \sum_{i=1}^N \pa_+ f_i\pa_- f_i
\bigr\} \>.
\eqn\seffthree
$$
The resulting equations of motion and constraint equations are
$$
\pa_+ \pa_- \chi = \pa_+ \pa_- \Omega
= -e^{{2 \over \kappa}(\chi - \Omega)} \>,
\eqn\sceomone
$$
$$
{1 \over \kappa}
[(\pa_{\pm} \Omega)^2 - (\pa_{\pm} \chi)^2]
+ \pa^2_{\pm} \chi + T_{\pm \pm}
- \kappa t_{\pm}  = 0 \>.
\eqn\scconone
$$

The field $\Omega$ can be viewed as a quantum corrected area
function.  At the horizon of a massive black hole it agrees to
leading order with the classical area function \areaone .  More
specifically, we will define the semi-classical area function as
$$
{\cal A} = \Omega - \Omega_{cr} \>,
\eqn\areatwo
$$
where
$\Omega_{cr}=\Omega(\gamma_{cr})
={\kappa\over 4}(1-\log{\kappa\over 4})$.
With this definition the area vanishes at the boundary curve.

The effective action \seffthree\ has a symmetry generated by the same
conserved current as we had in the classical theory,
$$
j^{\mu} = \nabla^{\mu} (\rho - \phi)
= {1 \over \kappa}\nabla^{\mu} (\chi - \Omega) \>.
$$
We can therefore again choose Kruskal coordinates, in which $\chi =
\Omega$, and the general solution of \sceomone\ takes the form
$$
\chi(\yp,\ym) = \Omega(\yp,\ym) =
\alpha_+(\yp) + \alpha_-(\ym) - \yp (\ym - P_{\infty}) \>,
\eqn\twofour
$$
where the functions $\alpha_{\pm}$ satisfy
$$
- \pa^2_{\pm} \alpha_{\pm} = T_{\pm \pm} - \kappa t_{\pm} \>.
\eqn\twofive
$$

\section{Semi-classical solutions}

It was observed in [\RSTtwo] that the global causal nature of
dynamical semi-classical geometries depends on the incoming energy
flux.  If the flux remains below a certain critical value,
$$
T_{++}(\sigma^+)<{\kappa\over 4} \>,
\eqn\cflux
$$
then no black hole is formed.  We will describe such low-energy
solutions later on.  Let us first focus on the case when the incoming
flux is above the critical value for some period of time,
$0<\sigma^+<\tau$.  The boundary curve then becomes space-like and
develops a curvature singularity.  A global event horizon, $H_G$,
separates the black hole region from the outside world.  The geometry
representing the black hole history in the semi-classical
approximation is shown in Figure~4.  The event horizon intersects the
space-like singularity at the endpoint of the evaporation process,
$(y^+_E,y^-_E)$.  The line segment $y^-=y^-_E$, $y^+<y^+_E$ is the
global event horizon.  The extension of this line to $y^+>y^+_E$ was
called the thunderpop in [\RSTtwo], and it divides the spacetime into
two regions called $I$ and $II$ as shown in Figure~4.  Region~$II$
represents the spacetime after the last bit of Hawking radiation has
gone past and therefore it is vacuum-like.\foot{The sharply defined
endpoint of the Hawking emission is presumably an artifact of the
semi-classical approximation in this model.  A more physical behavior
would be for the outgoing flux to die out gradually.  We will return
to this point in section~3.4.}  Region~$I$ covers the rest of the
spacetime.

\vskip 15pt
\vbox{
{\centerline{\epsfsize=3.0in \hskip 2cm \epsfbox{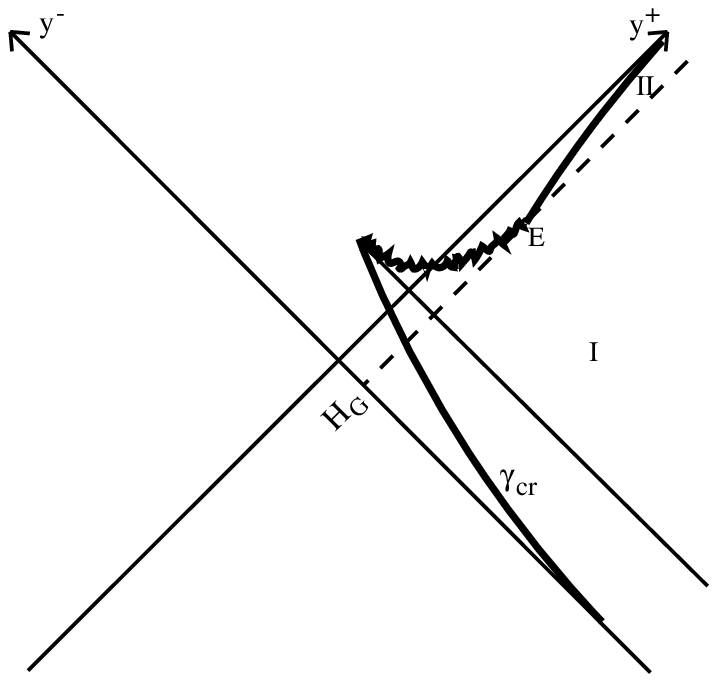}}}
\vskip 12pt
{\centerline{\tenrm FIGURE 4. Semi-classical black hole formation and
evaporation in Kruskal coordinates.}}
\vskip 15pt}

The solution in region~$I$ is
$$
\Omega = -y^+\bigl(y^-+P_+(y^+)-P_\infty\bigr) + M(y^+)
-{\kappa\over 4}\log{\bigl(-y^+(y^--P_\infty)\bigr)} \>,
\eqn\soli
$$
and in region~II it is the vacuum solution given by
$$
\Omega = -y^+y^- -{\kappa\over 4}\log{(-y^+y^-)} \>.
\eqn\solii
$$
Note that the global event horizon is not at $y^-=0$, as it was in
the classical case, but rather at
$$
y^-_E =
-{P_\infty \over e^{4M_\infty\over \kappa}-1} \>.
\eqn\yme
$$
For a large black hole mass, $y^-_E$ is exponentially close to zero.
The line $y^-=0$ still has special significance.  First of all it is
the asymptotic limit of the boundary curve $\gamma_{cr}$ as
$y^+\rightarrow\infty$.  Therefore it defines the boundary of the
region covered by the tortoise coordinates which are appropriate for
asymptotic observers.  Furthermore, if it were possible for signals
to propagate through the singularity along lines of constant $y^-$ to
reappear in the final vacuum-like region, then $y^-=0$ would indeed
be the global horizon.  We will call it the {\it ultimate horizon}.
At any rate, for massive black holes the values of $y^-$ at the
ultimate and global horizons are extremely close.

If the incoming energy flux remains below its critical value at all
times the boundary curve is everywhere timelike.  Semi-classical
solutions will have singularities there unless appropriate boundary
conditions are imposed [\RSTtwo,\RSTthree].  The curvature will be
finite at the timelike boundary if and only if
$$
\pa_+\Omega\big\vert_{\Omega=\Omega_{cr}}
=\pa_-\Omega\big\vert_{\Omega=\Omega_{cr}}
=0 \>.
\eqn\bconds
$$
These boundary conditions, along with the semi-classical equations of
motion, are sufficient to uniquely determine both the shape of the
boundary curve and the values of the semi-classical fields everywhere
in spacetime, for a given incoming energy flux.  We shall describe
some of these solutions in section~3.3.  Despite having some
attractive features these semi-classical solutions have some
unphysical properties.  This was part of our motivation to develop a
more physical picture in terms of a ``stretched horizon''.

\chapter{The stretched horizon}

Our postulates require us to build a theory in which a distant
observer makes no reference to events inside a black hole.  For this
purpose it is very useful to introduce the idea of a stretched
horizon, $H_S$, which is a visible timelike curve, in front of the
global event horizon of the black hole.  Each point on $H_S$ is
identified with a point on $H_G$, so the stretched horizon can act as
a ``surrogate'' for the global horizon in a phenomenological
description of black hole evolution.

\section{Definition and properties of the classical stretched
horizon}

We define the classical stretched horizon as follows.  Consider the
classical area function \areaone\ along the global event horizon.
For a black hole formed by gravitational collapse, depicted in
Figure~3, this area increases with $y^+$ until the black hole has
settled to its final size.  We define the stretched horizon by
mapping each point $m$ on the event horizon back along a
past-directed null line (away from the event horizon itself) to a
point $p$ at which
$$
{\cal A}(p) = {\cal A}(m) + \delta \>,
\eqn\hscl
$$
where $\delta$ is an arbitrary small constant.  This results in a
timelike curve as indicated in Figure~5.

\vskip 15pt
\vbox{
{\centerline{\epsfsize=3.0in \hskip 2cm \epsfbox{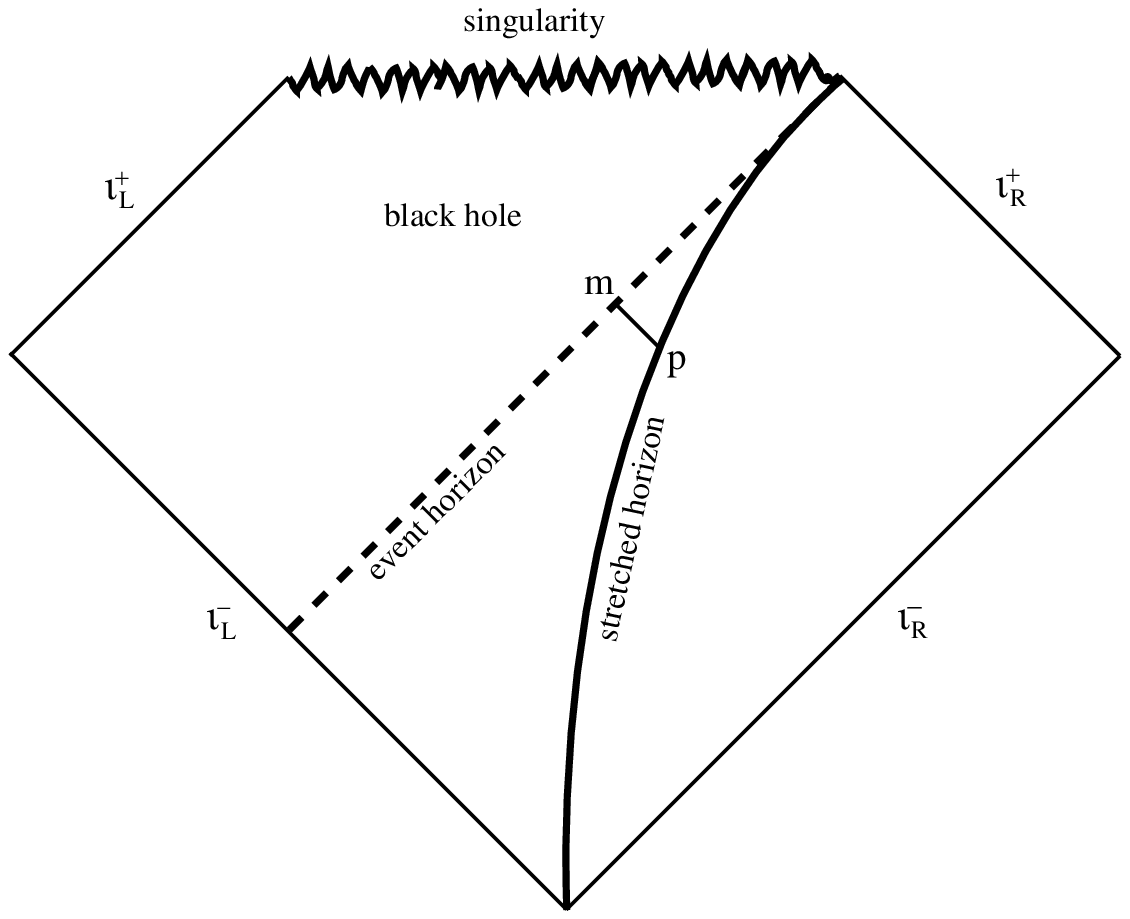}}}
\vskip 12pt
{\centerline{\tenrm FIGURE 5. Construction of the classical stretched
horizon.}}
\vskip 15pt}

Note that our definition differs from that given in [\Mempar].  There
the shift in the area between the global event horizon and the
stretched horizon scales like the horizon area itself.  For a massive
black hole our stretched horizon is thus much closer to the event
horizon.  Our definition is better suited for the semi-classical
theory of two-dimensional gravity considered here, and may also be
appropriate for the quantum description of black holes in four
spacetime dimensions.  For the classical black hole solutions \imone\
one finds the remarkably simple result that in Kruskal coordinates
the stretched horizon curve is independent of the incoming energy
flux, $T_{++}(y^+)$, and the curve is given by
$$
-y^+y^- = \delta \>.
\eqn\hsone
$$

The simplicity of the stretched horizon becomes even more apparent in
the tortoise coordinates \tortone .  While the event horizon lies at
$t=\infty$ and $\sigma =-\infty$, the stretched horizon is at fixed
spatial position
$$
\sigma_S = \half \log\delta \>.
\eqn\hstort
$$
Thus the stretched horizon can receive and emit signals.
Furthermore, to distant observers, clocks at the event horizon appear
infinitely slowed while they appear to run at a finite rate at the
stretched horizon.  For an eternal black hole of mass $M_0$, proper
time $\tau$ along $H_S$ is related to coordinate time $t$ by
$$
d\tau = \sqrt{\delta \over M_0{+}\delta}\, dt \>.
\eqn\threeone
$$

Finally, one can consider the Hawking temperature of a massive
two-dimensional dilaton black hole, $T = {1\over 2\pi}$, which is
independent of the mass.  Since temperature has units of energy, then
in proper time units at the stretched horizon the temperature is
$$
T_S = {1 \over 2\pi} {dt \over d\tau}
= {1 \over 2\pi} \sqrt{M_0{+}\delta \over \delta} \>.
\eqn\threetwo
$$
{}From \threetwo\ it appears that the local temperature at the
stretched horizon increases with $M_0$.  This is a bit misleading,
because the analogue of the Planck length in two-dimensional dilaton
gravity depends on the local value of the dilaton field, according to
$l_{pl} \sim e^\phi$.  At the stretched horizon, the dilaton field
satisfies $e^{-\phi}=\sqrt{M_0{+}\delta}$, so \threetwo\ implies
that, measured in Planck units, the temperature at the stretched
horizon is independent of the mass.  This result also holds for
four-dimensional black holes as we will see in section~5.1.

Let us continue examining the classical behavior of the stretched
horizon.  Consider the evolution of the area ${\cal A}$ on $H_S$.
Parametrizing $H_S$ by $\yp$ and substituting the definition \hsone\
of $H_S$ into the gravitational collapse solution \imone\ we find
$$
{\cal A}_S(\yp) = M(\yp) + \delta - \yp (P_+(\yp) - P_{\infty})
\eqn\threethree
$$
which, when differentiated twice, gives
$$
{d^2 {\cal A}_S \over (d\yp)^2} = -T_{++}(\yp) \>.
\eqn\threefour
$$
Transforming to tortoise coordinates, one can parametrize $H_S$ by
the tortoise time $t = \log\yp - \half \log\delta$.  Equation
\threefour\ then becomes
$$
{d^2 {\cal A}_S \over dt^2} - {d{\cal A}_S \over dt} = -\Bigl( {d\yp
\over dt} \Bigr)^2 T_{++}(\yp) \equiv -T_{++}(\sp)
\eqn\threesix
$$
where the quantity $T_{++}(\sp)$ is the incoming physical flux of
energy as seen by a distant observer.  There are two interesting
features of \threesix .  The first has to do with the nature of the
boundary conditions on the solutions of the equation.  In general,
the stretched horizon will begin to grow even before any energy
crosses it.  From \threesix\ we see that before $T_{++}$ becomes
nonzero, ${\cal A}_S$ has the solution
$$
{\cal A}_S(t) = Ce^t  \>.
\eqn\threeseven
$$
The choice of the constant $C$ is dictated by {\it final} conditions.
 As $t \rightarrow \infty$, a black hole is present with mass
$M_\infty$.  The area of the stretched horizon of such a black hole
is
$$
\lim_{t \to \infty} {\cal A}_S(t) = M_\infty + \delta \>,
\eqn\threeeight
$$
and thus \threeeight\ is the boundary condition one must impose on
the solution of \threesix .  This means that the initial state of the
stretched horizon must be tuned in conjunction with the incoming
matter distribution so that \threeeight\ is satisfied.  This strange
feature has been referred to in the membrane paradigm literature as
the ``teleological boundary condition'' [\Mempar].  We will show in
section~3.4 how the equations can, in fact, be given a more
conventional and causal interpretation.

The second interesting feature of \threesix\ is the dissipative term
${d{\cal A}_S \over dt}$.  It breaks time reversal symmetry much like
a friction term in ordinary mechanics.  The presence of dissipative
terms in mechanics is generally associated with the production of
heat and the increase of thermal entropy.  In the classical case, the
temperature of the black hole is zero, but in the semi-classical case
the temperature increases when the black hole is formed, and the
stretched horizon appears to radiate like a thermally excited black
body.

In the limit of large black holes, we can also consider the theory of
a massless matter field $f$ propagating in the black hole background.
 We find that the equations governing the matter fields interacting
with the stretched horizon also exhibit dissipation.  In the case of
a four-dimensional black hole interacting with electromagnetic
fields, the phenomenon of dissipation is described by attributing an
ohmic resistance to the
membrane.  A similar description can be given for two-dimensional
dilaton black holes.  Indeed, the theory of a massless field $f$
bears a useful resemblance to ordinary classical electrodynamics,
with $f$ playing the role of the vector potential.  We work in the
tortoise coordinates \tortone\ and define the ``electric'' and
``magnetic'' fields $E$ and $B$ by
$$
E = - \nabla_t f, \qquad B = \nabla_{\sigma} f \>.
\eqn\EandB
$$
The equation of motion for the $f$ field is
$$
\nabla^2 f = -4 \pi J
\eqn\feqn
$$
where we have introduced a source $j$.  Writing this equation in
terms of the fields $E$ and $B$, we obtain an inhomogeneous
``Maxwell'' equation
$$
\nabla_t E + \nabla_{\sigma} B
= -4 \pi J \>.
\eqn\Maxwellone
$$
We can also obtain the homogeneous ``Maxwell'' equation
$\nabla_{\sigma} E + \nabla_t B = 0$.

Now we consider the interaction of the $f$ field with the stretched
horizon, which, from the point of view of an external observer, is a
boundary absorbing all incoming waves.  This behavior can be modeled
by attributing a resistance to the stretched horizon,
$$
\nabla_t E + \nabla_{\sigma} B
= -4 \pi \rho_S^{-1}\, \delta (\sigma{-}\sigma_S)\, E \>.
\eqn\resist
$$
An incoming $f$ wave will be completely absorbed if and only if $\rho
= 4\pi$.  This is the analogue of the surface electrical resistivity
of a four-dimensional black hole.  The power absorbed by the
stretched horizon is $\rho_S^{-1} (\pa_t f)^2$, which can be thought
of as ohmic heating.  When quantum corrections are included, the heat
is radiated back as Hawking radiation.

\section{The semi-classical stretched horizon}

In defining the stretched horizon of a semi-classical black hole, we
find it more convenient to refer to the ultimate horizon at $\ym = 0$
than the event horizon \yme .  For a large black hole, the difference
is negligible.  We also replace the classical area function \areaone\
by its semi-classical counterpart \areatwo .  This leads to the
following condition for points on the stretched horizon:
$$
-\yp \ym - {\kappa \over 4} \log \Bigl( 1 -
{\ym \over P_{\infty}} \Bigr) = \delta \>,
\eqn\threethirteen
$$
where we have used the black hole solution \soli .  If the incoming
energy is large, then ${\ym \over P_{\infty}}$ will be very small on
$H_S$, except in the extremely early stages of its evolution.  Thus,
we will drop the log term in the definition.  In the classical case,
$\delta$ is an arbitrary small number.  In the semi-classical theory,
there is a natural choice, $\delta ={\kappa\over 4}$, for which the
area of the stretched horizon vanishes in the asymptotic past and
future when there is no black hole.  This implies that the stretched
horizon will coincide with the boundary curve $\gamma_{cr}$ in these
limits.  Thus, we define the stretched horizon to be the set of
points satisfying the condition
$$
- \yp \ym = {\kappa\over 4} \>.
\eqn\hstwo
$$
In tortoise coordinates, the stretched horizon is given by the curve
$$
{\hat \sigma}_S (t)
= \half \log \Bigl({\kappa \over 4} \Bigr)
= \sigma_S  \>.
\eqn\hsthree
$$

Let us now consider the black hole evolution in tortoise coordinates
as shown in Figure~6.  The incoming flux is assumed to be vanishing
outside the interval $0<\sp<\tau$.  For $\yp < 1$, \ie\ $\sp < 0$, we
have the initial linear dilaton vacuum and the boundary curve is
given by
$$
{\hat \sigma}_{cr} (t) = \log \Bigl( {P_{\infty}
e^t \over 2} \Bigl[ \sqrt{1 + {\kappa e^{-2t}
\over P^2_{\infty}}} - 1 \Bigr] \Bigr) \>.
\eqn\threefourteen
$$
In the remote past, this curve tends to
$$
{\hat \sigma}_{cr} (t) \rightarrow \sigma_S
- {P_{\infty} \over \sqrt{\kappa}} e^t \>.
\eqn\threefifteen
$$
We see that the boundary begins to separate from the stretched
horizon at a time
$$
t^* = -\log P_{\infty} + \half \log \kappa \>.
\eqn\threesixteen
$$
The incoming matter arrives at the stretched horizon at a time $t_0
=-\sigma_S$.  Consequently, we see a period of time $\sim \log
P_{\infty}$ during which the boundary moves in anticipation of the
infalling matter.  It continues to move toward $\sigma = -\infty$
with a velocity which approaches that of light.

A second boundary curve passes through the naked singularity at the
endpoint of the black hole evaporation.  Behind the stretched horizon
the second boundary curve is space-like and coincides with the
curvature singularity.  This is shown in Figure~6.  The
semi-classical viewpoint is that the infalling matter becomes trapped
between these boundary lines and disappears into a spatially
disconnected region.  However, our postulates do not require us to
pay any attention to this region, as it lies behind the stretched
horizon.

\vskip 15pt
\vbox{
{\centerline{\epsfsize=3.0in \hskip 2cm \epsfbox{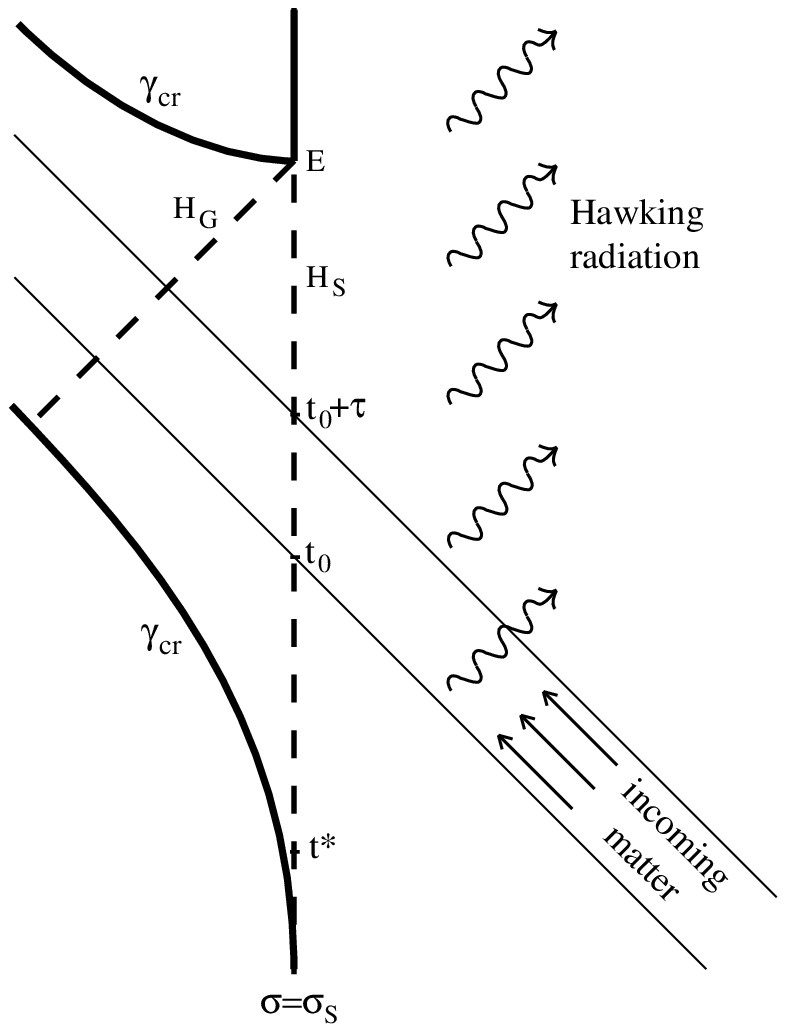}}}
\vskip 12pt
{\centerline{\tenrm FIGURE 6. Black hole evolution in tortoise
coordinates.}}
\vskip 15pt}

Now let us turn to the outgoing Hawking radiation.  Using the $(--)$
constraint equation \scconone , one finds the outgoing Hawking flux
$$
T_{--}(t,\sigma) = {\kappa \over 4} \Bigl[ 1
- {1 \over (1 - P_{\infty} e^{(t-\sigma)})^2} \Bigr]
\Theta (t_E - t + \sigma - \sigma_S)\>,
\eqn\threeeighteen
$$
where
$$\eqalign{
t_E &= \log{(e^{{4\over \kappa}M_\infty}-1)}
-\log{P_\infty} + \sigma_S \cr
&\approx {4\over \kappa}M_\infty -
\log{P_\infty} + \sigma_S  \>.\cr}
\eqn\endtime
$$
The outgoing flux has its leading, albeit somewhat fuzzy, edge along
the null curve
$$
\sigma - t \approx \log(P_{\infty}) \>.
\eqn\threetwenty
$$
Tracing this line back to the stretched horizon we find that it
intersects the stretched horizon at the time
$$
t^* = t_0 - \log{P_\infty} \>.
\eqn\threetwentyone
$$
If we interpret the outgoing thermal radiation as originating on the
stretched horizon, it begins at a time well before the incoming
matter arrives.  For early times, the semiclassical area of the
stretched horizon is approximately
$$
{\cal A}_S \approx \half P^2_{\infty} e^{2t} \>.
\eqn\threetwentytwo
$$
The radiation begins {\it just as the area of the stretched horizon
begins to increase.}  The radiation has turned on by the time the
area (and entropy) of the stretched horizon have increased to their
values at $t^*$, given by ${\cal A}^* \approx {\kappa \over 8}$.

The correspondence between the onset of Hawking radiation and the
excitation of the stretched horizon is unexpected.  From a strictly
local point of view, nothing special is happening at this point.

It is straightforward to generalize the equation \threesix\ governing
the evolution of the area of the stretched horizon.  Using \soli\ and
\hstwo\ and parametrizing $H_S$ by $\yp$, we find
$$
{\cal A}_S(\yp) = M(\yp) - \yp[P_+(\yp)
- P_{\infty}] + {\kappa \over 4} (1 - \log
({\kappa\over 4} + \yp P_{\infty})) \>.
\eqn\threetwentythree
$$
Differentiating twice with respect to $\yp$ and transforming to
tortoise coordinates gives
$$
{d^2 {\cal A}_S \over dt^2} - {d {\cal A}_S \over dt}
= -T_{++}(t,\sigma_S)
+ {\kappa \over 4} \Bigl[ {1 \over (1 + {\kappa
\over 4 P_{\infty}}
\exp [-(\sigma_S{-}t)])^2} \Bigr] \>.
\eqn\threetwentyfour
$$
Once the stretched horizon area is significantly greater than
${\kappa \over 4}$, the second term on the right hand side can be
simplified to ${\kappa \over 4}$, giving
$$
{d^2 {\cal A}_S \over dt^2} - {d {\cal A}_S \over dt}
= -T_{++}(t,\sigma_S)
+ {\kappa \over 4} \>.
\eqn\threetwentyfive
$$
The second term on the right hand side of \threetwentyfive\
represents the effects of the outgoing Hawking radiation on the
evolution of ${\cal A}_S$.  For example, we see that a stationary
solution is possible if $T_{++} (\sigma^+) = {\kappa \over 4}$.  In
this case, the incident energy flux is just sufficient to balance the
outgoing thermal radiation.  In section~4, we will see that things
are somewhat more complicated, and that ${\cal A}_S$ has a brownian
motion superimposed on its average motion.

Let us now review the process of formation and evaporation as seen by
a distant observer using tortoise coordinates.  The infalling matter
is scheduled to begin passing the stretched horizon at time $t_0$.
However, well before this, at time $t_0 - \log P_{\infty}$, the
stretched horizon begins to separate from the boundary, and its area
increases by an amount of order ${\kappa \over 8}$.  Assuming the
standard connection between entropy and area, this is the point at
which the stretched horizon becomes thermally excited.  The distant
observer sees the onset of Hawking radiation originating from this
point.  At the time $t_0$, the infalling matter is swallowed behind
the stretched horizon, which continues to radiate.  If we assume that
there are microphysical degrees of freedom which underlie the
thermodynamic description, $t_0$ is the first opportunity for them to
feel the infalling matter.  Therefore, at least for the initial time
of order $\log P_{\infty}$, no information can be stored in the
Hawking radiation [\svv].

After the infalling matter is absorbed, the area begins to decrease.
The acceleration term in \threetwentyfive\ goes to zero, and ${\cal
A}_S$ satisfies
$$
{d{\cal A}_S \over dt} = -{\kappa \over 4} \>.
\eqn\threetwentysix
$$
The area, entropy, and mass of the black hole tend linearly to zero.
The entire process from $t_0$ to the endpoint at which ${\cal A}_S$
returns to its initial value takes a time $t \approx {4M \over
\kappa}$, during which a constant flux of Hawking radiation is
emitted by the stretched horizon.  As we shall see in section~4, the
entire semi-classical evolution is accompanied by random brownian
fluctuations, which introduce an uncertainty of order $\sqrt{M}$ to
the lifetime of the process.

\section{Incident flux below the black hole threshold}

We now consider the case in which the incident energy flux remains
below the critical value $\kappa\over 4$ for all time.  The resulting
geometry and outgoing flux of radiation was obtained in [\RSTthree].
Here we will transcribe some of those results into tortoise
coordinates.  Assume that the incoming energy flux $T_{++}(\sp)$
vanishes outside the interval $0{<}\sp{<}\tau$.  The time-like
boundary curve $\gamma_{cr}$ is obtained by solving the equations of
motion \sceomone\ subject to the RST boundary conditions \bconds .
In tortoise coordinates, $\gamma_{cr}$ satisfies the following
equation:
$$
e^{{\hat \sigma}_{cr}(t)} -
{\kappa\over 4} e^{-{\hat \sigma}_{cr}(t)}
= -\int_{{\hat \sigma}_{cr}(t)}^{\infty} ds e^{-s} T_{++}(t+s) \>.
\eqn\bcurve
$$
\vbox{
\vskip -24pt
{\centerline{\epsfsize=3.0in \hskip 2cm \epsfbox{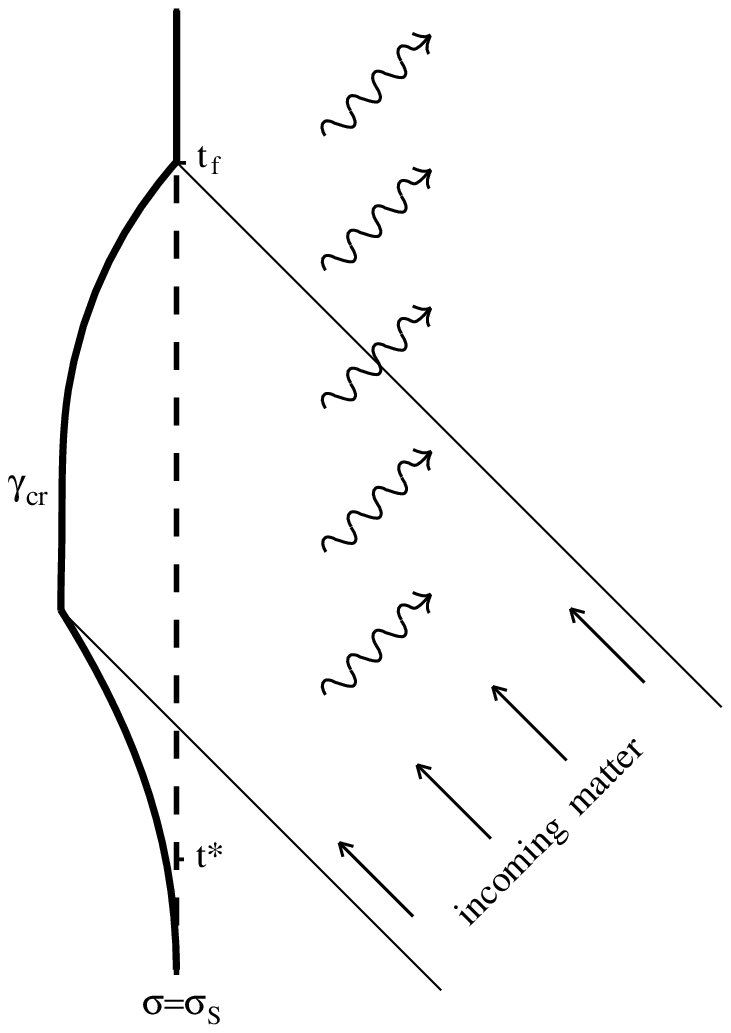}}}
\vskip 12pt
{\centerline{\tenrm FIGURE 7. Sub-critical flux of incident matter.}}
\vskip 15pt}
As in the black hole case, in the remote past the boundary curve
tends to a fixed spatial position,
$$
\sigma_S = \half \log \Bigl({\kappa\over 4}\Bigr) \>,
\eqn\sigmas
$$
which we will continue to call the location of the stretched horizon,
even though no black hole is formed.  The boundary begins to move
exponentially in anticipation of the incoming matter, as it did in
the black hole case, but this time $\gamma_{cr}$ remains time-like
throughout the evolution, and eventually returns to the stretched
horizon at the time $t_f = \tau - \sigma_S$.  This is shown in
Figure~7.

The outgoing flux is obtained by applying the following reflection
conditions [\RSTthree]:
$$
T_{--}(\sigma^-)-{\kappa\over 4}
= \bigl({d\hat\sigma^+_{cr}\over d\sigma^-}\bigr)^2
\,\bigl[
T_{++}(\hat\sigma^+_{cr})-{\kappa\over 4}\bigr] \>,
\eqn\refl
$$
where $\hat\sigma^+_{cr}(\sigma^-)$ denotes the boundary curve
$\gamma_{cr}$ parametrized by $\sigma^-$.  Note that this
prescription for reflecting energy flux does not involve boundary
conditions imposed directly on the matter fields.

It is instructive to consider a constant incoming energy flux,
$T_{++}(\sigma^+)={\overline T}$, of duration $\tau$.  The early time
behavior of the boundary curve is given by
$$
\hat\sigma_{cr}(t)
= \sigma_S - {P_\infty\over \sqrt{\kappa}}\,e^t \>.
\eqn\rise
$$
The boundary curve continues to recede from the stretched horizon
until the incoming flux intersects $\gamma_{cr}$, as shown in
Figure~7.  If ${\overline T}$ is much smaller than ${\kappa\over 4}$,
the boundary curve never moves appreciably away from the stretched
horizon, but if ${\overline T}$ is close to ${\kappa\over 4}$,
$\gamma_{cr}$ moves deep into the region of negative $\sigma$.  The
maximum coordinate distance between $\gamma_{cr}$ and $H_S$ is
$$
\sigma_{max} = \half \log{\bigl(
{\kappa\over 4}-{\overline T}(1{-}e^{-\tau})
\bigr)} \>,
\eqn\sigmax
$$
which occurs at time $t_{max}=-\sigma_{max}$.  After that the
boundary curve begins to return to the stretched horizon.  How fast
it returns depends on the parameters ${\overline T}$ and $\tau$.  In
the limit of very long duration of the incoming energy flux, $\tau
\gg 1$, the boundary remains practically stationary for a long time
at its maximum distance, but eventually it returns and arrives back
at the stretched horizon at time $t_f=\tau-\sigma_S$.  If, on the
other hand, the duration of the incoming energy flux is relatively
short the boundary curve rapidly returns and approaches $H_S$ with a
velocity
$$
v\approx {{\overline T}\over {\kappa\over 2}-{\overline T}} \>.
\eqn\speed
$$
When the incoming energy flux goes to the critical value,
${\kappa\over 4}$, this velocity approaches the speed of light.

In [\RSTthree], it was speculated that the critical boundary might
behave like a moving mirror reflecting the $f_i$ fields.  We can now
see that this can only be consistent in the limit of small incoming
energy flux, ${\overline T} \ll {\kappa\over 4}$.  For if ${\overline
T} \approx {\kappa\over 4}$, the incoming radiation would be met by a
very relativistic mirror, which would greatly blueshift the reflected
radiation.  In addition, accelerated mirrors create incoherent
quantum radiation of net positive energy.  The result would be far
more energy output than the total incoming energy.  It is therefore
clear that only a tiny fraction of the incident energy can be
coherently reflected by the boundary when ${\overline T} \approx
{\kappa\over 4}$.

The outgoing flux of energy can be calculated from \refl\ and one
finds that almost all the energy is radiated back before the incoming
signal could have been reflected from the boundary curve
$\gamma_{cr}$.  The following odd rule gives a better account of the
energy output as determined by the RST boundary conditions in the
case ${\overline T} \approx {\kappa\over 4}$: assume that at time
$t^* = -\log{P_\infty\over \sqrt{\kappa}}$, when the boundary curve
$\gamma_{cr}$ separates from $H_S$, the stretched horizon becomes
thermally excited to a temperature $T\approx {1\over 2\pi}$.  Assume
that the hot horizon emits thermal radiation at a fixed rate until
time $t_f$.  The total radiated energy will be
$$
E_{out}\approx {\kappa \over 4}(t_f-t^*) \>,
\eqn\eout
$$
which accounts for about the right amount of energy output.  We will
provide further motivation for this alternate viewpoint in the
following subsection.

Another interesting point concerns the fate of the conserved charges
associated with the global $O(N)$ symmetry of the matter fields.
Only when the energy flux ${\overline T}$ is much less than ${\kappa
\over 4}$ can the boundary curve consistently behave like a mirror,
reflecting both the energy and the conserved charges.  When
${\overline T} \approx {\kappa \over 4}$, almost all the energy is
radiated before the charges have an opportunity to reflect.  If a
large total charge of order $(t_f{-}t^*)$ came in and was reflected,
it would have to be carried by a small energy, of order $\kappa$.  In
other words, it would have to be carried out in the form of quanta
with energy of order ${\kappa\over (t_f{-}t^*)}$ which would take a
very long time.  It is easy to see, however, that as ${\overline T}$
approaches ${\kappa \over 4}$, the reflected region does not spread
out as it would have to if it were composed of quanta of ever lower
energy.  Therefore, only a small amount of conserved charge can be
reflected.  In addition to thermalizing the incident energy, the
process must also destroy the conservation of quantum numbers.

Another failure of the boundary to behave like a mirror can be
illustrated by considering an interruption in an otherwise uniform
incoming flux -- a glitch.  If the boundary behaved like a mirror, a
brief, sharp interruption would be expected in $T_{--}$ where the
glitch reflects off the boundary, but an explicit calculation shows
that this is not the case.

\section{A causal description of the stretched horizon}

Our aim is a self-contained description of black hole evolution as
seen by a distant observer in which no reference need be made to
events behind the stretched horizon.  It will also become clear that
such a formulation has significant advantages in the low energy
sector, compared to a semi-classical description which focuses on
boundary conditions imposed at the boundary curve $\gamma_{cr}$.  In
particular, the stretched horizon offers a unified view, in which it
is no longer necessary to treat the cases of large and small incident
energy flux separately.

We saw earlier that the stretched horizon begins to expand in what
appears to be a teleological manner before the incoming matter
arrives.  One might be concerned that this would preclude a
conventional causal Hamiltonian description of the quantum stretched
horizon.  We do not believe this to be the case.  From a formal point
of view, the cause of the horizon expansion is a gravitational
dressing which is attached to the incoming energy flux.

Consider the initial state description in tortoise coordinates.
Suppose an incoming flux of energy is described by $T_{++}(\sp)$.
The functions $P_+$ and $M$ are given in tortoise coordinates by
$$
\eqalign{P_+(\sp) &= \int_{-\infty}^{\sp} du\, e^{-u}
\,T_{++}(u) \>, \cr
M(\sp) &= \int_{-\infty}^{\sp} du \,T_{++} (u) \>, \cr}
\eqn\masstwo
$$
and the field $\Omega$ is given by
$$
\Omega = e^{2\sigma} + \bigl[ P_{\infty} - P_+(\sp)
\bigr] e^{\sp} + M(\sp) - {\kappa \over 4} \bigl[ \sp
+ \log (P_{\infty} + e^{-\sm}) \bigr]
\eqn\threetwentyseven
$$
Let us subtract from $\Omega$ the functional form $\overline \Omega$
which it would have in the absence of any incoming matter,
$$
{\overline \Omega} = e^{2\sigma} - {\kappa \over 2} \sigma \>.
\eqn\threetwentyeight
$$
We obtain
$$
\omega \equiv \Omega - {\overline \Omega}
= \omega_{\inn}(\sp) + \omega_{\out}(\sm) \>,
\eqn\threetwentynine
$$
where
$$\eqalign{
\omega_{\inn}(\sp) &= \bigl(P_{\infty}
- P_+(\sp) \bigr) e^{\sp} + M(\sp) \>, \cr
\omega_{\out}(\sm) &= - {\kappa \over 4} \log \bigl( 1
+ P_{\infty} e^{\sm} \bigr) \>. \cr}
\eqn\threethirty
$$
We see that the free field $\omega$ consists of an incoming part and
an outgoing part.  We can use the outgoing $\omega$ field to
determine the outgoing energy-momentum flux.

Since the incoming part is completely determined by the incoming
energy flux, we will consider it to be a ``dressing'' of the incoming
matter.  It can be written
$$
\omega_{\inn}(\sp) = \int du\, T_{++}(u)\, W(\sp-u) \>,
\eqn\threethirtyone
$$
where
$$
W(\sp-u) = \Theta(u-\sp)e^{(\sp-u)} + \Theta(\sp-u) \>.
\eqn\threethirtytwo
$$
In other words, a bit of energy $\delta M$ arriving along the curve
$\sp = u$ must be accompanied by an $\omega$ dressing which has the
value $W(\sp-u)\>\delta M$.  The $\omega$ dressing precedes the
incoming $f_i$ flux, and is the first thing that strikes the
stretched horizon.  By time reversal symmetry, a bit of outgoing
energy $\delta M$ departing along the curve $\sm = v$ also has an
$\omega$ dressing given by $W(v-\sm)\>\delta M$.

In a complete quantum theory, the outgoing state would be described
by a vector in the physical state space of $f_i$-particles, from
which it would be possible to compute the expectation value of
$T_{--}$.  This would not be feasible, however, even if we knew the
exact nature of the microstructure of the stretched horizon.  As we
shall now see, thermodynamic arguments can give information about
$\omega_{\out}$, which is sufficient to compute $T_{--}$.

The $\omega$ dressing of the incoming matter satisfies
$$
( \pa_+ - \pa^2_+ )\, \omega_{\inn}(\sp) = T_{++} \>.
\eqn\threethirtyfour
$$
By time reversal we obtain a relation between $\omega_{\out}$ and
$T_{--}$:
$$
(\pa_- + \pa^2_-)\, \omega_{\out}(\sm) = -T_{--} \>.
\eqn\threethirtyfive
$$
This can be written as a condition at the stretched horizon,
$$
{d\omega_{\out} \over dt} =
-T_{--}\bigl(t, {\hat \sigma}_S(t)\bigr)
- {d^2 \omega_{\out} \over dt^2}
\eqn\threethirtysix
$$
The outgoing thermal flux $T_{--}$ is assumed to originate at the
thermally excited stretched horizon.  The entropy of the stretched
horizon is given by $\omega = \omega_{\inn} + \omega_{\out}$, since
at the stretched horizon, $\omega = {\cal A}$.

In thermal equilibrium, $T_{--}$ should be a well-defined function of
the thermodynamic state of the stretched horizon, and thus of the
entropy $\omega_S$.  More generally, for non-equilibrium processes
such as the onset and end of Hawking evaporation, the flux may depend
on the detailed time history.  Nevertheless, we will consider a
simplified model in which $T_{--}$ depends only on the instantaneous
value of $\omega_S$, and the sign of its time derivative.  In other
words, we shall allow for the possibility that $T_{--}(\omega)$ has
different functional forms at the beginning and end of the
evaporation.  Thus, we assume the radiated flux is a function of
$\omega$,
$$
T_{--} = T_{--}(\omega) = T_{--}(\omega_{\inn}
+ \omega_{\out}) \>.
\eqn\threethirtyseven
$$
Substituting this into \threethirtysix\ gives
$$
{\dot \omega}_{\out} + {\ddot \omega}_{\out}
= -T_{--}(\omega_{\out} + \omega_{\inn}) \>.
\eqn\threethirtyeight
$$
Since $\omega_{\inn}(t)$ is known in terms of $T_{++}$, we have
obtained a differential equation for $\omega_{\out}$.

To obtain agreement with the semi-classical theory for large black
holes, we assume that $T_{--}(\omega)$ approaches the value ${\kappa
\over 4}$ for $\omega \gg \kappa$.  Of course, $T_{--}(\omega)$
should be zero for $\omega = 0$, which is the ground state of the
stretched horizon.  Furthermore, if one requires that the onset of
the Hawking radiation, for a massive black hole, agrees with the
semi-classical result \threeeighteen , then one is led to a specific
form for $T_{--}(\omega)$ during the heating phase at early times.
This can be constructed as follows.  Define a function $z(\omega)$ by
$$
\omega = z(\omega) - {\kappa \over 4} \log \Bigl( 1
+ {4 \over \kappa}z(\omega) \Bigr) \>,
\eqn\threethirtynine
$$
then choose
$$
T_{--}(\omega) = {\kappa \over 4} \Bigl[ 1 - {1
\over (1 + {4\over \kappa}z(\omega))^2} \Bigr] \>.
\eqn\threefourty
$$
For massive black holes, it can be shown that $\omega(t)$ as given by
\threetwentynine\ solves \threethirtyeight\ until near the end of the
evaporation.

We can now see an important advantage of the stretched horizon
formulation, concerning the endpoint of Hawking evaporation.  In the
semi-classical RST model [\RSTtwo], overall energy conservation could
only be achieved by having a negative energy ``thunderpop'' at the
end of the evaporation process.  The negative energy was bounded and
small, but nevertheless an embarrassment [\past].  In contrast,
consider the sum of \threethirtyfour\ and \threethirtyfive ,
evaluated at the stretched horizon,
$$
{\dot \omega}_{\inn} - {\ddot \omega}_{\inn}
+{\dot \omega}_{\out} + {\ddot \omega}_{\out}
= T_{++} - T_{--} \>.
\eqn\threefourtyone
$$
Integrating both sides of \threefourtyone\ over time reveals that
energy is conserved, \ie ,
$$
\int dt \> T_{++} = \int dt \> T_{--}
\eqn\threefourtytwo
$$
as long as $\omega$ begins {\it and\/} ends at zero.  This is assured
in the remote past, when the stretched horizon coincides with the
boundary curve.  For the late time evolution of $\omega$, we return
to \threethirtyeight .  From \threethirty\ we see that the late value
of $\omega_{\inn}$ is the total infalling mass, so $\omega_{\inn}
\rightarrow M_{\infty}$.  Inserting this into \threethirtyeight\
leads to the following differential equation for $\omega_{\out}$ at
late time:
$$
{\ddot \omega}_{\out} + {\dot \omega}_{\out}
+ T_{--}(\omega_{\out} + M_{\infty}) = 0 \>.
\eqn\threefourtythree
$$
This is the equation for the damped motion of a particle subject to a
restoring force, with equilibrium position at $\omega_{\out} +
M_{\infty} = 0$.  Provided the motion is overdamped, we find that
$\omega$, and therefore $T_{--}$, tend smoothly to zero at late
times.  This places a condition on $T_{--}(\omega)$, namely that it
goes to zero no slower than ${\omega \over 4}$.  Comparing with
\threethirtynine\ and \threefourty\ we see that $T_{--}$ must depend
differently on $\omega$ during the cooling and heating phases.

%
%
\chapter{Brownian motion of the horizon\foot{{\rm The material in
this section is based on work done in collaboration with N.~Seiberg,
S.~Shenker, and J.~Tuttle [\ssstt].}}}

Lagrangian mechanics and thermodynamics are quite different
descriptions of a system.  According to the usual principles of
lagrangian mechanics, the motion of any system is reversible and the
concepts of heat and entropy have no place.  Thermodynamics, on the
other hand, is the theory of the irreversible dissipation of
organized energy into heat.  The thermodynamic description arises
from the coarse graining of the mechanical description, in which
configurations which are macroscopically similar are considered
identical.

The equations of semi-classical gravity are peculiarly thermodynamic
near the stretched horizon.  In this section, we will see that they
include another effect that generally occurs in thermodynamic
systems, namely, random fluctuation and diffusion.  That such an
effect should occur was pointed out to us by N. Seiberg and S.
Shenker [\seishe].  Specifically, we shall see that the area of a
two-dimensional dilaton black hole undergoes brownian motion and
diffuses away from its semi-classical value.  This phenomenon can be
independently understood from thermodynamics and quantum field
theory.

We begin by recalling the Einstein relation between specific heat and
energy fluctuations.  The average energy and squared energy of a
system in thermal equilibrium with a heat bath are
$$
\vev{E} = - {1 \over Z} {\pa Z \over \pa \beta} \>,
\eqn\fourone
$$
$$
\vev{E^2} = {1 \over Z}{\pa^2 Z \over \pa \beta^2}
\eqn\fourtwo
$$
where $Z$ is the partition function and $\beta$ is the inverse
temperature.  From \fourone\ and \fourtwo\ it follows that
$$
{\pa \vev{E} \over \pa \beta} = \bigl[ \vev{E}^2
- \vev{E^2} \bigr] = -\Var(E) \>,
\eqn\fourthree
$$
where $\Var(X) = \big \langle (X - \vev{X})^2 \big \rangle$ denotes
the variance of the quantity $X$.  This can be expressed in terms of
the specific heat $C$, defined by
$$
C = {\pa \vev{E} \over \pa T} =
-{1 \over T^2} {\pa \vev{E} \over \pa \beta} \>,
\eqn\fourfour
$$
so that
$$
T^2 C = \hbox{Var}(E) \>.
\eqn\fourfive
$$
In particular, since the variance of any quantity is
positive-definite, the specific heat is also positive definite.  When
applied to a four-dimensional Schwarzschild black hole, \fourfour\
gives nonsense because the specific heat is negative.  This is a sign
of instability.

In section~3.2, we obtained a dynamical equation \threetwentyfive\
for the time dependence of the horizon area ${\cal A}$.  In thermal
equilibrium, the incoming and outgoing average energy fluxes are both
equal to ${\kappa \over 4}$.  In this case \threetwentyfive\ has a
static solution for each value of the average area.  Since the
Hawking temperature of two-dimensional black holes is independent of
the mass in the semi-classical approximation, the specific heat of a
black hole is infinite.  By \fourfive , the root-mean-square
fluctuations of the mass, and therefore the area, are also infinite.
This means that the thermal fluctuations will so smear the horizon
that the different mass static black hole solutions should be
replaced by a single ensemble for all masses and areas.

More generally, a time-dependent semi-classical black hole will have
a brownian motion superimposed on the semi-classical solution.  Among
other effects, this will cause a statistical fluctuation in the
elapsed time before the black hole ceases to radiate.  The
fluctuation will be of order $\sqrt{M}$, where $M$ is the initial
black hole mass.

Physically, we can understand this as follows: fluctuations in the
thermal flux of energy at the horizon cause the black hole mass to
randomly increase and decrease with time.  For an ordinary system,
with positive specific heat, such fluctuations are self-regulating.
A momentary increase (decrease) in the energy of the system causes an
increase (decrease) in its temperature, which in turn causes heat to
flow back to (from) the reservoir, thus restoring equilibrium.  In
the present case, the temperature does not respond to the energy
fluctuation.  Therefore, there is no tendency to return to the
original energy balance.  The mass and area just random walk away
from their original values.

If a black hole of area ${\cal A}_0$ is created at time $t = 0$ and
is subsequently illuminated with thermal radiation at the Hawking
temperature $T = {1 \over 2\pi}$, then at time $t > 0$ the mass of
the black hole will have random walked:
$$
\bigl\langle ({\cal A}(t) - {\cal A}_0)^2
\bigr\rangle \propto t \>.
\eqn\foursix
$$
The exact coefficient in \foursix\ can be computed from a knowledge
of fluctuations in the thermal energy of the matter fields in the
surrounding bath.  For the case of $N$ massless fields, the result is
$$
\bigl\langle ({\cal A}(t) - {\cal A}_0)^2
\bigr\rangle = {N \over 24\pi^2} t \>.
\eqn\fourseven
$$
A rough translation of this result into Kruskal coordinates can be
made by observing that $y^{\pm}$ are exponentials of tortoise
coordinates.  Equation \fourseven\ suggests that in terms of an
infrared cutoff in Kruskal coordinates $\log R \approx t$, the
fluctuations in the horizon area satisfy
$$
\hbox{Var}({\cal A}) \approx {N \over 24\pi^2} \log R \>.
\eqn\foureight
$$

Now let us consider the semi-classical field equations \sceomone .
In particular, the scalar field $\Omega$ satisfies an inhomogeneous
free field equation in Kruskal coordinates given by
$$
\pa_+ \pa_- \Omega = -1 \>.
\eqn\fournine
$$
The static black hole solutions to \fournine\ have the form
$$
\Omega_M = M - \yp \ym
\eqn\fourten
$$
and the semi-classical area of the event horizon of a massive black
hole is given by
$$
{\cal A} \approx \Omega_M(0) = M \>.
\eqn\foureleven
$$
The area of the stretched horizon is a bit larger but this difference
will not be important in this section.

Now let us consider the quantum fluctuations about \fourten .  Define
$\Delta = \Omega - \Omega_M$.  The fluctuation $\Delta$ satisfies a
free wave equation,
$$
\pa_+ \pa_- \Delta = 0 \>.
\eqn\fourtwelve
$$
This suggests that $\Delta$ is a canonical, massless free field.  As
such it has fluctuations which are logarithmically infrared
divergent,
$$
\bigl\langle \Delta^2(0) \bigr\rangle
\approx {\kappa \over 2\pi^2} \, \log R \>,
\eqn\fourthirteen
$$
where $R$ is the Kruskal coordinate infrared cutoff.  This estimate
of the fluctuations in the horizon area precisely agrees with the
thermodynamic result \fourseven .  It should be pointed out that
there are technical subtleties involved in the quantization of this
model, and the above result has not been rigorously established.
However, the agreement with thermodynamics strongly suggests that
$\Delta$ behaves like a canonical field [\ssstt].

%
%
\chapter{Consequences of the postulates}

\section{Microstructure of the stretched horizon}

Consider a quantum field theory in a two-dimensional spacetime with a
strictly time-like boundary.  Suppose the boundary is stationary at
$\sigma = 0$, except for a brief time interval $[t_a, t_b]$, during
which it moves toward negative $\sigma$ (left) and then returns.  The
fields are defined to the right of the boundary.  The boundary may
have additional degrees of freedom.

Without loss of generality we can pretend that the boundary is
permanently at $\sigma = 0$ by assigning it extra degrees of freedom
during the interval $[t_a, t_b]$.  During this period the system has
field degrees of freedom on the negative $\sigma$ axis.  Nothing
prevents us from formally considering these degrees of freedom to
belong to the boundary at $\sigma=0$.

In the case of subcritical flux, where the boundary is always
time-like and in causal contact with distant observers, we can
perform a similar formal trick, regardless of the nature of the
boundary degrees of freedom.  They, as well as the fields behind the
stretched horizon, can be formally assigned to the stretched horizon.
 We gain nothing from this except the assurance that a set of
stretched horizon degrees of freedom can be defined.  Note that this
procedure in no way influences the experiences of an observer
crossing the stretched horizon.

Up to now we have assumed nothing radical.  The fact that outside
observers see an apparently real stretched horizon is surprising but
derivable from conventional semi-classical assumptions.  At this
point we will make a radical departure from traditional thought about
black holes, which is required by our three postulates.  We propose
that, for the purposes of a distant observer,

{\it A consistent set of quantum mechanical degrees of freedom
continue to describe the stretched horizon even when the critical
flux is exceeded.}

We postulate no details about these degrees of freedom, but some
general properties are required by Postulate 3.  According to
standard thermodynamic reasoning, the entropy of a large system is
the negative of the logarithm of the density of states.  For both
two- and four-dimensional black holes, the entropy is proportional to
the area.  We therefore require that the dimension of the Hilbert
space of a stretched horizon with area ${\cal A}$ is of order
$\exp({\cal A})$.  For a four-dimensional black hole, this suggests
that the number of degrees of freedom per unit area is a universal,
intensive property, independent of the total mass of the black hole.

This universality of stretched horizon properties is general.  Define
the stretched horizon of a four-dimensional Schwarzschild black hole
to have an area {\it one Planck unit greater\/} than the global event
horizon.  The local rate of clocks at the stretched horizon is easy
to compute.  The analogue of \threeone\ has the form
$$
{d\tau \over dt} \sim {M_P \over M}
\eqn\fiveone
$$
where $M_P$ is the Planck mass and $M$ is the mass of the black hole.
 The local proper temperature at the stretched horizon, $T_S$, is
related to the asymptotically measured Hawking temperature $T_H$ by
$$
T_S = {M \over M_P} T_H \>.
\eqn\fivetwo
$$
Using the standard Hawking temperature $T_H \sim {M_P^2\over M}$
gives the universal value
$$
T_S \sim M_P \>.
\eqn\fivethree
$$
The total energy of the black hole, measured in proper units at the
stretched horizon, is
$$
M_S = M {dt \over d\tau} \sim {M^2 \over M_P} \>.
\eqn\fivefour
$$
Dividing this by the area of the stretched horizon we find the
surface energy density to be
$$
{M_S \over {\cal A}} \sim M_P^3 \>.
\eqn\fivefive
$$

In the semi-classical theory defined in Section~3, the temperature of
a black hole is completely independent of its mass.  Thus, as a black
hole evaporates, its energy flux is exactly constant, until the
instant it disappears.  An immediate consequence of Postulate~3 is
that the temperature of a two-dimensional dilaton black hole cannot
be strictly constant when the mass tends to zero.  However, there can
be a maximum temperature, which is quickly saturated as energy
increases.  Suppose there are discrete energy levels with a density
$\rho(E)$ which behaves asymptotically as
$$
\rho(E) \sim \exp(2\pi E) \quad
\hbox{as} \quad E \rightarrow \infty \>.
\eqn\fivesix
$$
The partition function
$$
Z(\beta) = \sum_{\hbox{\sevenrm states}} e^{-\beta E}
\eqn\fiveseven
$$
converges for all $\beta > 2\pi$.  For large $\beta$, $Z$ can be
approximated by the first few terms:
$$
Z \approx 1 + e^{-\beta E_1} + \ldots \>,
\eqn\fiveeight
$$
and the average energy is given by
$$
\langle E \rangle = - {\pa \log (Z) \over \pa \beta}
\approx E_1\, e^{-\beta E_1} + \ldots \>.
\eqn\fivenine
$$
As the $\langle E \rangle$ tends to infinity, the temperature tends
to the value $T = {1\over 2\pi}$ in agreement with the semi-classical
limit.  As the energy tends to zero, the temperature $T ={1\over
\beta}$ also tends to zero.

In the semi-classical approximation, the black hole radiates a bit
more energy than the system originally had [\RSTtwo].  This was
compensated by a final ``thunderpop'' of negative energy.  From the
present point of view, a more plausible behavior is that as the black
hole nears the endpoint of the evaporation process, its temperature
and luminosity tend to zero and do not overshoot.  Note that this is
precisely the behavior exhibited by solutions of \threefourtythree .

Another unphysical consequence of the semi-classical theory concerns
static solutions, corresponding to a uniform sub-critical energy
flux, as the limit of critical flux is approached.  According to the
semi-classical theory a static solution exists for every value of the
energy flux, $T_{++}=T_{--}={\overline T}<{\kappa\over 4}$.  The
semi-classical area \areatwo , evaluated at the stretched horizon
\hstwo , is given by
$$
{\cal A}_S = {\overline T}
+({\kappa\over 4}-{\overline T})\bigl[
\log{({\kappa\over 4}-{\overline T})}
-\log{\kappa\over 4}\bigr] \>.
\eqn\sharea
$$
As ${\overline T}$ goes to ${\kappa\over 4}$ the area approaches
${\cal A}={\kappa\over 4}$.  On the other hand, the statistical
theory of the previous section requires the mean square value of the
area to diverge in this limit.  The model in section~3.4, based on
the thermodynamics of the stretched horizon, does exhibit that
behavior.  In general, thermodynamic boundary conditions at the
stretched horizon, as in section~3.4, yield a more consistent
physical description than semi-classical boundary conditions imposed
at the critical curve, $\gamma_{cr}$.

The large horizon fluctuations as the temperature approaches
$T={1\over 2\pi}$ are reminiscent of critical
behavior.\foot{R.~Laughlin has also suggested similarities between
black hole behavior and phase transitions [\bobl].}  As the critical
temperature is approached the area of the stretched horizon
fluctuates more and more, until the horizon swallows up all of space.
 In the case of a second order phase transition, correlated domains
fluctuate and grow until one domain swallows up the whole sample.
The failure of semi-classical theory to correctly account for the
horizon fluctuations is analogous to the failure of mean field theory
in critical phenomena.

\section{Thermal entropy {\it vs.} entropy of entanglement}

Given our assumptions about the microstructure of the stretched
horizon, it is evident that no real loss of information takes place
during black hole evaporation.  Nevertheless, it is far from clear
how the large amount of initial data is stored in outgoing
thermalized radiation.  Our discussion of this subject will follow
the very illuminating study by Don Page [\page].

Let us begin by distinguishing two kinds of entropy.  The first,
which is of purely quantum origin, we call {\it entropy of
entanglement.}  Consider a quantum system composed of two parts, $A$
and $B$.  In what follows, $B$ will refer to the stretched horizon
and $A$ to the radiation field outside the stretched horizon.  Assume
the Hilbert space of state vectors ${\cal H}$ is a tensor product
space: ${\cal H} = {\cal H}_A \otimes {\cal H}_B$.  If $\{ \ket{a}
\}$ is an orthonormal basis for ${\cal H}_A$ and $\{ \ket{b} \}$ is
an orthonormal basis for ${\cal H}_B$, then a general ket
$\ket{\psi}$ in ${\cal H}$ may be written
$$
\ket{\psi} = \sum_{a, \, b} \psi (a, b) \ket{a} \otimes \ket{b} \>.
\eqn\fiveten
$$
The density matrix of the subsystem $A$, in the basis $\{ \ket{a}
\}$, is
$$
\rho_A(a, a') = \sum_{b} \psi(a, b) \psi^*(a', b) \>,
\eqn\fiveeleven
$$
and that of $B$ is
$$
\rho_B(b, b') = \sum_{a} \psi(a, b) \psi^*(a, b') \>.
\eqn\fivetwelve
$$
Note that the composite system $A \cup B$ is in a pure state.

The entropies of entanglement of subsystems $A$ and $B$ are defined
by
$$
\eqalign{S_E(A) &= -\hbox{Tr}\bigl(\rho_A \hbox{LOG}(\rho_A) \bigr)
\>, \cr
S_E(B) &= -\hbox{Tr}\bigl(\rho_B \hbox{LOG}(\rho_B) \bigr) \>. \cr}
\eqn\fivethirteen
$$
It is easy to prove that $S_E(A) = S_E(B)$ if the composite system is
in a pure state.  The entropy of entanglement of a subsystem is only
zero if $| \psi \rangle$ is an uncorrelated product state.  The
entropy of entanglement is {\it not\/} the entropy with which the
second law of thermodynamics is concerned; $S_E$ can increase or
decrease with time.  A final point is that if the dimension of ${\cal
H}_B$ is $D_B$, then the maximum value of $S_E(B)$ (and therefore of
$S_E(A)$) is
$$
S_E(B)_{\hbox{\sevenrm max}} = - \log (D_B) \>.
\eqn\fivefourteen
$$
We have assumed in \fivefourteen\ that $D_B \leq D_A$.

The second kind of entropy is {\it entropy of ignorance.}  Sometimes
we assign a density matrix to a system, not because it is quantum
entangled with a second system, but because we are ignorant about its
state, and we assign a probability to each state.  For example, if we
know nothing about a system, we assign it a density matrix
proportional to the unit matrix.  If we know only its energy, we
assign a density matrix which is vanishing everywhere except the
allowed energy eigenspace.  Thermal entropy is of this type: it
arises because of practical inability to follow the fine grained
details of a system.  For a system in thermal equilibrium with a
reservoir, we assign a Maxwell-Boltzmann density matrix
$$
\rho_{\hbox{\sevenrm MB}} = Z^{-1} \hbox{EXP}(-\beta H) \>,
\eqn\fivefifteen
$$
and define the thermal entropy by
$$
S_T = -\hbox{Tr}\bigl(\rho_{\hbox{\sevenrm MB}}\,
\hbox{LOG}(\rho_{\hbox{\sevenrm MB}}) \bigr) \>.
\eqn\fivesixteen
$$

Now let us consider the evolution of both kinds of entropy during the
formation and evaporation of a two-dimensional black hole.  Let us
begin with the thermal entropy of the stretched horizon, which we
assume is equal to its area.  As we have seen, the area begins to
increase exponentially with $t$ before the infalling matter arrives,
reaching its maximum at the arrival time.  The area then decreases
linearly with $t$ until the black hole disappears.  This is
illustrated in Figure~8.  The thermal entropy of the outgoing
radiation begins to increase due to the emission from the excited
stretched horizon.  Shortly after the radiation begins, the
temperature approaches $T={1\over 2\pi}$, so that the rate of change
of the thermal entropy of the radiation is constant throughout most
of the process.  This is also shown in Figure~8.

\vskip 15pt
\vbox{
{\centerline{\epsfsize=3.0in \hskip 2cm \epsfbox{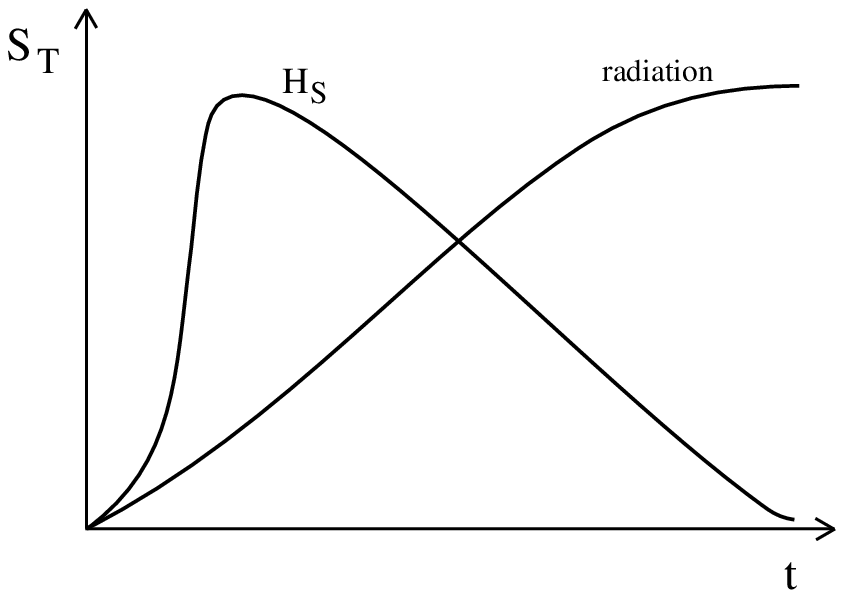}}}
\vskip 12pt
{\centerline{\tenrm FIGURE 8. Thermal entropy of stretched horizon
and radiation field as a function of time.}}
\vskip 15pt}

Now consider the entropy of entanglement.  Initially, the stretched
horizon is in its ground state, with minimal area, and the radiation
field is described by a pure state.  The entropy of entanglement
starts at zero.  As soon as the stretched horizon area begins to
increase, $f-$quanta are emitted.  Typically, the state of the
$f-$quanta will be correlated to the state of the stretched horizon,
so that $S_E$ will start to increase.  However, $S_E$ will generally
be bounded by the logarithm of the dimension of the Hilbert space
describing the stretched horizon, which we have assumed is
proportional to the area.  In other words, at any time,
$$
S_E(H_S) \leq S_T(H_S) = {\cal A}(t) \>.
\eqn\fiveseventeen
$$
Thus, the entropy of entanglement is bounded and must return to zero
as the area of the stretched horizon returns to its vacuum value.
Page has argued that in the beginning the entropy of entanglement is
likely to approximately follow the thermal entropy of the radiation
field, so that the history of $S_E$ should look like Figure~9.

\vskip 15pt
\vbox{
{\centerline{\epsfsize=3.0in \hskip 2cm \epsfbox{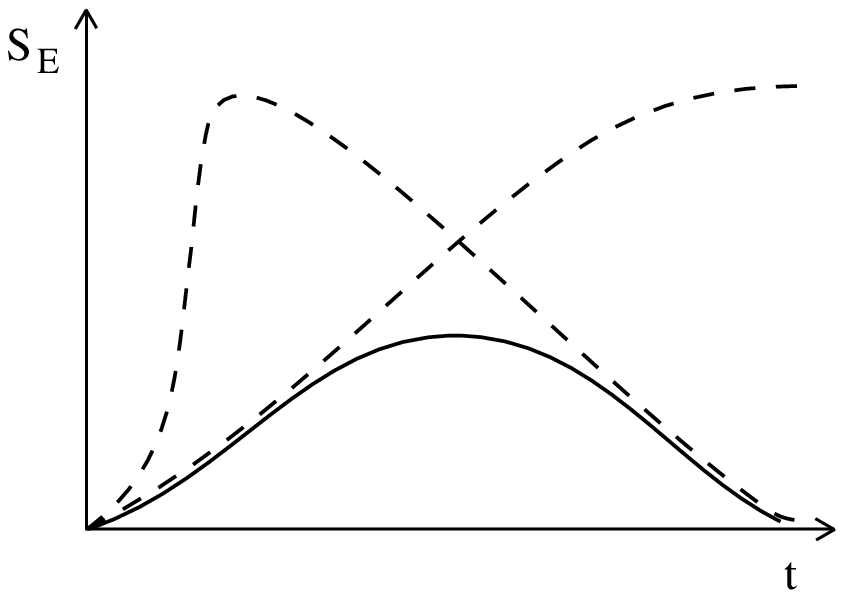}}}
\vskip -40pt
{\centerline{\tenrm FIGURE 9. Entanglement entropy of radiation and
stretched horizon.}}
{\centerline{\tenrm The dashed curves indicate the thermal entropies
of Figure~8.}}
\vskip 15pt}

Evidently, as time elapses, subtle differences develop between the
coarse grained thermal density matrix of the radiation and the exact
description.  Postulates 1 and 3 dictate that the entropy of
entanglement return to zero in a more or less definite way as the
black hole evaporates.  In particular, there is no room for a stable
or very long-lived remnant storing the incident information.

To understand the difference between the thermal and exact density
matrices of the final outgoing radiation, consider a time about
half-way through the evaporation process, when the thermal entropy
and the entropy of entanglement are still not too different.  The
total fine grained entropy of the combined system of stretched
horizon and radiation is zero, but the radiation is correlated to the
degrees of freedom of the stretched horizon.  More time elapses, and
the stretched horizon emits more quanta.  The previous correlations
between the stretched horizon and the radiation field are now
replaced by correlations between the early part of the radiation and
the newly emitted quanta.  In other words, the features of the exact
radiation state which allow $S_E$ to return to zero are {\it long
time correlations spread over the entire time occupied by the
outgoing flux of energy.}  The local properties of the radiation are
expected to be thermal.  For example, the average energy density,
short time radiation field correlations, and similar quantities that
play an imporant role in the semi-classical dynamics should be
thermal.  The long time correlations which restore the entropy to
zero are not important to average coarse grained behavior, and are
just the features which are not found unless a suitable microphysical
description is provided for the stretched horizon.

\section{Discussion}

We will conclude with some speculation about the nature of the
stretched horizon microstructure for four-dimensional black holes.
If we consider nearly spherical black holes, a stretched horizon can
be defined as follows.  Consider a radial incoming null geodesic
which crosses the global horizon where its area is ${\cal A}$.
Proceed backward along such geodesics until the surface with area one
Planck unit larger is encountered.  By using such ingoing geodesics,
we can map every point of the global horizon to a point on the
stretched horizon.

The global horizon is composed of a bundle of light rays which can be
thought of as a two dimensional fluid on the global horizon
[\Mempar].  The points of this fluid can be mapped to the stretched
horizon, thereby defining a fluid flow on that surface.  Classically,
the fluid behaves as a continuous, viscous fluid with conventional
shear viscosity and negative bulk viscosity.  A natural candidate for
the microphysics of the stretched horizon is to replace the
continuous classical fluid with a fluid of discrete ``atoms''.

As we have seen, the intensive thermodynamic variables of the
stretched horizon are universal and do not depend on the size or mass
of the black hole.  This demands that the surface density of atoms
also is independent of the area.  When incoming energy flux or
outgoing Hawking radiation causes the area of a patch of the
stretched horizon to change, points of the fluid will pop into and
out of existence in order to keep the density constant.

Finally we would like to point to a feature of 3+1-dimensional black
holes which is not shared by the 1+1-dimensional theory.  This
feature adds plausibility to the claim that the stretched horizon is
in thermal equilibrium
during most of the evaporation.  Consider an observer at the
stretched horizon who counts the number of particles emitted per unit
proper time.  Since the stretched horizon is always at the Planck
temperature the number of particles emitted per unit area per unit
proper time is of order one in Planck units.  If all these particles
made it out to infinity, then a distant observer would estimate a
number of particles emitted per unit time, which is obtained by
multiplying by the black hole area and the time dilation factor,
$$
{dN\over dt}\sim  M^2 {d\tau\over dt} \sim M \>.
\eqn\parnum
$$
On the other hand, the number per unit time of particles that
actually emerge to infinity is obtained by multiplying the black hole
luminosity $L\sim {1\over M^2}$ by the inverse energy of a typical
thermal particle at the Hawking temperature.  The result is
$$
{dN\over dt}\sim {1\over M} \>.
\eqn\parnumm
$$
Therefore it seems that most of the particles emitted from the
stretched horizon do not get to infinity.  In fact, only those
particles which are emitted with essentially zero angular momentum
can overcome the gravitational attraction of the black hole, and the
rest fall back [\Mempar].  This gives rise to a thermal atmosphere
above the stretched horizon which only slowly evaporates and whose
repeated interaction with the stretched horizon insures thermal
equilibrium.  Such a thermal atmosphere can be obtained in $1+1$
dimensions by including massive degrees of freedom in the model, and
this may indeed be necessary for a fully consistent description of
two-dimensional black hole evaporation.

If the considerations of this paper are correct then black holes
catalyze a very different phenomenon than that envisioned by Hawking
[\haw].  To begin with an incoming pure state of matter composed of
low-energy particles falls into its own gravitational well.  The
matter is blue-shifted relative to stationary observers so that when
it arrives at the stretched horizon it has planckian wavelengths.
Thereupon it interacts with the ``atoms'' of the stretched horizon
leading to an approximately thermal state.  The subsequent
evaporation yields approximately thermal radiation but with
non-thermal long time correlations.  These non-thermal effects depend
not only on the incoming pure state but also on the precise nature of
the Planck-scale ``atoms'' and their interaction with the
blue-shifted matter.  The evaporation products then climb out of the
gravitational well and are red-shifted to low energy.  The result is
remarkable.  The very low-energy Hawking radiation from a massive
black hole has non-thermal correlations, which contain detailed
information about Planck-scale physics [\tHooftone,\svv].  The
phenomenon is reminiscent of the imprinting of planckian fluctuations
onto the microwave background radiation by inflation.

The view of black holes that we have presented is, of course,
incomplete.  As we have emphasized, the reality of the membrane can
not be an invariant which all observers agree upon.  Furthermore,
although conventional quantum field theory in an evaporating black
hole background seems to lead to a description in which a single
state vector describes the interior and exterior of the black hole,
this description must be wrong if our postulates are correct.
Precisely what is wrong is not clear to us, but we wish to emphasize
that the event space for an experiment should only contain physically
measurable results.

In many respects, the situation seems comparable to that of the early
part of the century.  The contradictions between the wave and
particle theories of light seemed irreconcilable, but careful thought
could not reveal any logical contradiction.  Experiments of one kind
or the other revealed either particle or wave behavior, but not both.
 We suspect that the present situation is similar.  An experiment of
one kind will detect a quantum membrane, while an experiment of
another kind will not.  However, no possibility exists for any
observer to know the results of both.  Information involving the
results of these two kinds of experiments should be viewed as {\it
complementary \/} in the sense of Bohr.

\ack
We are very grateful to a number of our friends for collaboration,
advice and encouragement.  In particular we have benefitted from many
conversations with T.~Banks, A.~Bilal, C.~Callan, S.~de~Alwis,
S.~Giddings, S.~Hawking, R.~Laughlin, D.~Page, J.~Preskill,
A.~Strominger, E.~Verlinde, and H.~Verlinde.  Section~4 of the paper
is based on work done in collaboration with N.~Seiberg, S.~Shenker
and J.~Tuttle.  Much of the remainder of the paper reflects our
previous close collaboration with J.~Russo.  Finally, we are
especially thankful to G.~'t~Hooft for inspiration and for
reinforcing our own belief in the inviolability of the principles of
quantum theory in black hole evaporation.  This work is supported in
part by National Science Foundation grant PHY89-17438.  J.~U. is
supported in part by a National Science Foundation Graduate
Fellowship.

\refout
\end